\documentclass[aps,pre,twocolumn,10pt,groupedaddress,amsfonts,amssymb,amsmath,showpacs,showkeys,a4paper,floatfix]{revtex4-1}

\usepackage[english]{babel}
\usepackage{blindtext}
\usepackage{graphicx}
\usepackage{bm}
\usepackage{multirow}
\usepackage{mathrsfs}

\usepackage[usenames,dvipsnames,svgnames,table]{xcolor}

\definecolor{a_color}{HTML}{377EB8}
\definecolor{b_color}{HTML}{4DAF4A}
\definecolor{c_color}{HTML}{984EA3}
\definecolor{d_color}{HTML}{FF7F00}
\definecolor{e_color}{HTML}{E41A1C}
\definecolor{f_color}{rgb}{0,0,0}

\newcommand{\aclr}[1]{{\color{a_color}#1}}
\newcommand{\bclr}[1]{{\color{b_color}#1}}
\newcommand{\cclr}[1]{{\color{c_color}#1}}

\newcommand{\fclr}[1]{{\color{f_color}#1}}

\newcommand{\integral}[1]{\left\langle{#1}\right\rangle}

\newcommand*\Laplace{\mathop{}\!\mathbin\bigtriangleup}

\usepackage{listings}

\lstdefinestyle{customc}{
  belowcaptionskip=1\baselineskip,
  breaklines=true,
  frame=TB,
  xleftmargin=\parindent,
  language=C,
  showstringspaces=false,
  escapechar=@
}
\begin{document}

\title{Entropic multi-relaxation lattice Boltzmann scheme for turbulent flows}

\author{Fabian B\"osch}
\email{boesch@lav.mavt.ethz.ch}
\affiliation{Department of Mechanical and Process Engineering, ETH Zurich, 8092 Zurich, Switzerland}

\author{Shyam S. Chikatamarla}
\affiliation{Department of Mechanical and Process Engineering, ETH Zurich, 8092 Zurich, Switzerland}

\author{Ilya Karlin}
\affiliation{Department of Mechanical and Process Engineering, ETH Zurich, 8092 Zurich, Switzerland}

\date{\today}

\begin{abstract}
We present three dimensional realizations of the model introduced recently by (Karlin, B\"osch, Chikatamarla, Phys. Rev. E 2014; Ref~\cite{KBC}) and review the role of the entropic stabilizer. The presented models achieve outstanding numerical stability in presence of turbulent high Reynolds number flows. We report accurate results for low order moments for homogeneous isotropic decaying turbulence and second order grid convergence for most assessed statistical quantities. The explicit and efficient nature of the scheme renders it a very promising candidate for both engineering and scientific purposes in the vicinity of highly turbulent flows.
\end{abstract}


\maketitle

\section{Introduction}

The lattice Boltzmann method (LB)  \cite{Benzi1992145,succi} is a modern and highly successful kinetic-theory approach to 
computational fluid dynamics (CFD) and computational physics of complex flows and fluids, with applications ranging from
turbulence \cite{TURBO} to flows at a micron scale \cite{Ansumali2007} and multiphase flows  \cite{Swift1995,AliPRL}, 
to relativistic hydrodynamics \cite{Miller1}, soft-glassy systems \cite{Benzi2009} and beyond. 

While conventional methods for CFD attempt to solve the Navier-Stokes equations, LB's 
underlying equations form a kinetic system which can be thought of as a discrete analogue to Boltzmann's equation. Carefully designed 
populations $f_i(\bm{x},t)$ corresponding to a set of discrete velocity vectors $\bm{v}_i$, $i=1,\dots,b$ spanning a regularly spaced lattice 
with nodes $\bm{x}$. Conceptually, the dynamics of populations $f_i$ can be split into free flight (advection) and collison (relaxation)
which is reflected by the general one-parametric LB equation 
\begin{equation}\label{eq:Egeneral}
f_i(\bm{x}+\bm{v}_i,t+1)=f_i'\equiv(1-\beta)f_i(\bm{x},t)+\beta f_i^{\rm mirr}(\bm{x},t).
\end{equation}
Here the left-hand side is the propagation of the populations along the lattice links, while the right-hand side is the so-called 
{post-collision} state $f'$. The relaxation parameter $\beta$ is associated with the transport coefficient of the macroscopic target 
equation (the incompressible Navier-Stokes equations herein). The mirror state $f^{\rm mirr}$ represents the maximally over-relaxed state.
Realization of hydrodynamics  was made possible, in the first place, with the lattice 
Bhatnagar-Gross-Krook (LBGK) model \cite[]{Chen92,Qian92}, in which one takes
\begin{equation}\label{eq:LBGKmirr}
f_i^{\rm mirr}=2f_i^{\rm eq}-f_i.
\end{equation}
Here $f_i^{\rm eq}$ is the equilibrium which is found as a maximizer of the entropy \cite{ansumaliMinimal},
\begin{equation}
\label{eq:S}
S[f]=-\sum_{i=1}^b f_i\ln\left(\frac{f_i}{W_i}\right),
\end{equation}
subject to fixed locally conserved fields identified by the first $D+1$ velocity moments, $\rho=\sum_{i=1}^b f_i$ (density) 
and $\rho\bm{u}=\sum_{i=1}^b \bm{v}_i f_i$ (momentum density), 
and where the weights $W_i$ are lattice-specific constants. 
With the proper symmetry of the lattice, the LBGK equation, \eqref{eq:Egeneral} and \eqref{eq:LBGKmirr}, recovers the Navier-Stokes 
equation for the fluid velocity $\bm{u}$, with the kinematic viscosity
\begin{equation}\label{eq:viscosity}
\nu=c_s^2\left(\frac{1}{2\beta}-\frac{1}{2}\right),
\end{equation}
where $c_{\rm s}$ is the speed of sound (a lattice dependent $O(1)$ constant). 
%
The LBGK model is unambiguous since $\beta  \in (0,1)$ is fixed by the kinematic viscosity \eqref{eq:viscosity}.  
The limit $\beta\to 1$ (small kinematic viscosity) is particularly important as it is pertinent to achieving, if only in principle, 
high Reynolds number regimes. 
A notable departure from the continuous BGK approximation of Boltzmann's equation becomes manifest in the feature of over-relaxing towards
the mirror-state which disconnects LBGK from the kinetic theory domain $\beta \in (0,\tfrac{1}{2}]$ \cite{KarlinExactLB}. 


Almost immediately after its inception, the LBGK model has taken lead in the lattice Boltzmann approach to the
simulation of complex hydrodynamic phenomena \cite[]{succi,Aidun10}, and remains the ``working horse'' of the LB methods to-date. 
Popularity of LBGK is primarily based on its simplicity and exceptional computational efficiency.
Despite of its promising nature and popularity the LBGK model shows severe deficiencies (disruptive numerical instabilities) 
already at moderate Reynolds numbers. This precluded the LB method from making a sustainable impact in the field of 
computational fluid dynamics. 

A number of approaches can be found in the literature intended to alleviate this issue. We will restrict the following short 
discussion to methods without explicit turbulence models. Most notably, the entropic lattice Boltzmann method (ELBM) features 
non-linear stability and has shown excellent performance \cite{Karlin99,cylinder,CK_PRL1}. 
While ELBM converges to LBGK in the resolved case, it locally alters the relaxation parameter which in turn modifies the viscosity 
in order to fulfil the second law of thermodynamics by both enhancing and smoothing the features of the flow where necessary, 
subject to an entropy condition.

The dimension $b$ of the kinetic space populated by $f_i$ is usually greater than strictly necessary for recovering the Navier-Stokes 
equations. In three dimensions, for instance, ten linearly independent basis vectors would suffice to represent the conserved variables and
the symmetric stress tensor $\Pi_{\alpha\beta} = \sum_{i=1}^b v_{i\alpha} v_{i\beta} f_i$. Although it can be shown that the coupling 
to the non-hydrodynamic higher ("ghost") moments cannot be chosen arbitrarily in the limit ${\rm Ma} \rightarrow 0$  \cite{Dellar2003351} 
independent relaxation of these moments may have a favourable effect on the numerical stability.

The multiple relaxation parameters (MRT) LB schemes follow this line of thinking. While the relaxation of the off-diagonal parts of the stress tensor are fixed by the 
choice of kinematic viscosity, the MRT scheme embraces that the relaxation of higher order moments should not affect the dynamics of 
the flow field (up to the Navier-Stokes level) and hence can be used to construct more stable LB schemes. Based on a separation of scales
between fast and slowly varying moments several MRT schemes  were suggested for the choice of relaxation of higher order moments (beyond the stress tensor) 
\cite{HSB,dHumieres92,PhysRevE.61.6546,Geier2006}. The choice of relaxation parameters is crucial in order to increase the operational 
range in terms of stability and requires careful tuning.

In order to eliminate the influence of higher order moments which may be rapidly oscillating and cause numerical instabilities
the regularized LB scheme (RLB) \cite{Chen2006125,Latt2006} was proposed. While the relaxation parameters for MRT schemes are tunable, they are fixed for RLB such that the higher order non-equilibrium moments vanish. As any finite lattice representation introduces 
discrete artifacts among the higher order moment tensors, the regularization operation ensures isotropy, albeit in the confined subspace limited 
up to the stress tensor level. 
Although MRT and RLB models were successful 
in slightly stabilizing the LB method, they still remain challenged by high Reynolds numbers \cite{Freitas2011}. 

Recently, the authors have developed a scheme without the need for tunable parameters or turbulent viscosity 
(Karlin, B\"osch, Chikatamarla, Phys. Rev. E 2014; Ref\cite{KBC})
which has demonstrated a significant extension in the operation range for simulations at high Reynolds numbers. 
Promising results have been reported for both two and three dimensions, as well as for complex boundaries and in 
presence of turbulence. Much like ELBM, entropic considerations have been employed to render the scheme stable 
without introducing considerable computational overhead and by keeping the simplicity and locality of the LBGK and MRT schemes. 
Below we shall refer to this class of models as KBC models for brevity.

In the remainder of this manuscript we present the construction of a family of KBC models for the standard D3Q27 lattice and review the details of the entropic stabilization. The Kida vortex flow and a randomly generated initial condition shall be employed as turbulent benchmark flows and are studied in detail for one of the realizations of KBC and the stability domain of LBGK, RLB and KBC variations is assessed numerically.

\section{Equilibrium on the standard lattice in three dimensions}

In this section, we shall review the standard lattice in three dimensions and the corresponding equilibrium. While the material presented in this section is of a review character, we shall highlight some fundamental features of the equilibrium which were not fully discussed so far in the literature.

The standard lattice in a dimension $D$ is built as tensor (direct) product of $D$ copies of the fundamental one-dimensional set  $v_{(i)}=i$, where $i=0,\pm 1$:
\begin{equation}\label{eq:D1Q3}
v_{-1}=-1,\ v_0=0,\ v_1=1.
\end{equation}
The natural $D$-dimensional Cartesian reference frame generated by the tensor product of $D$ copies of the fundamental set (\ref{eq:D1Q3}) makes it convenient to enumerate the discrete velocities accordingly. Considering the three-dimensional case $D=3$ below, we write for any of the $b=27$ discrete velocities, 
\begin{equation}
\label{eq:D3Q27vel}
\bm{v}_i=(v_{ix},v_{iy},v_{iz}),\ i=1,\dots,27;\ v_{i\alpha}\in\{-1,0,1\}.
\end{equation}

The equilibrium on this standard product-lattice maximizes the entropy \eqref{eq:S} subject to fixed conservation laws of density and momentum \cite{ansumaliMinimal}. It is written most elegantly in the following product-form:

\begin{equation}
\label{eq:entropic_eq}
f_{i}^{eq} =  \rho W_i\Psi\left[ B(u_x)\right]^{v_{ix}}
\left[B(u_y)\right]^{v_{iy}}\left[B(u_z)\right]^{v_{iz}},
\end{equation}
where the weights $W_i$ are
\begin{equation}\label{eq:weights}
W_i=W_{v_{ix}}W_{v_{iy}}W_{v_{iz}},
\end{equation}
and function $\Psi$ is universal for all the discrete velocities (it does not depend on the discrete velocity index),
\begin{equation}
\Psi(\bm{u})= A(u_x)A(u_y)A(u_z),
\end{equation}
with
\begin{equation}
A(u)= 2 - \sqrt{1+3 u^2}.
\end{equation}
Furthermore, the universal function $B(u)$ contributing to the formation of the last term in (\ref{eq:entropic_eq}) is written as
\begin{equation}
B(u)= \frac{2 u+ \sqrt{1+3 u^2}}{1-u}.
\end{equation}
Finally, the weights $W_{\alpha}$ in (\ref{eq:weights}) are dictated by the speed of sound,
\begin{equation}\label{eq:sound}
c_s=\frac{1}{\sqrt{3}},
\end{equation}
and are
\begin{equation}
W_{0}=2/3,\ W_{-1}=1/6,\ W_{1}=1/6.
\end{equation}


Remind that the Maxwellian in $D$ dimensions is written as a product with respect to an arbitrarily fixed Cartesian reference frame, in accord with the familiar property of the shifted Gaussian distribution,
\begin{equation}\label{eq:Maxwell}
e^{-\frac{(\bm{v}-\bm{u})^2}{T}}=\left(\prod_{\alpha=1}^D e^{-\frac{v_{\alpha}^2}{T}}\right)
\left(\prod_{\beta=1}^D e^{-\frac{u_{\beta}^2}{T}}\right)
\left(\prod_{\gamma=1}^D \left[e^{\frac{2u_{\gamma}}{T}}\right]^{v_{\gamma}}\right).
\end{equation}
It is easy to recognize the Maxwellian character of the product-lattice equilibrium. The multiplication of the weights in (\ref{eq:entropic_eq}) corresponds to the first multiplier, the function $\Psi$ corresponds to the second multiplier, while the product of functions $B$ reflects the last multiplier.



However, the true Maxwellian is isotropic, as also reflected by reading equation (\ref{eq:Maxwell}) from right to left, the products collapse to a dependence on the kinetic energy in the co-moving frame alone, and the reference to the arbitrarily fixed Cartesian coordinates disappears.  This is not so with the discrete velocities. It is imperative therefore to demonstrate that the product-form (\ref{eq:entropic_eq}) is {\it manifestly isotropic} to the order of accuracy of the lattice Boltzmann model. This can be done most elegantly in the following way: Instead of expanding each population (\ref{eq:entropic_eq}) into powers of the velocity components $u_{\alpha}$, let us first expand the logarithm of $f_i^{\rm eq}$ (we consider a generic case of $D$ below):
\begin{equation}
\ln f_i^{\rm eq}=\ln\rho+\ln W_i+\sum_{\alpha=1}^{D}\ln A(u_{\alpha})+\sum_{\alpha=1}^{D}v_{i\alpha}\ln B(u_{\alpha}).
\end{equation}
Let us denote $[\varphi(u)]_2$ the operation of the second-order truncation of the expansion of any function $\varphi$ around $u=0$ to get

\begin{equation}
\left[\ln f_i^{\rm eq}\right]_2=\ln\rho+\ln W_i+\left(1-\frac{3}{2}(\bm{u}\cdot\bm{u})\right)+3(\bm{v}_i\cdot\bm{u}),
\end{equation}
where we have used the standard notation,
\[\bm{a}\cdot\bm{b}=\sum_{\alpha=1}^Da_{\alpha}b_{\alpha},\]
for the Cartesian scalar product of $D$-dimensional vectors.
Then, using the identity, $
[f_i^{\rm eq}]_2=\left[\exp([\ln f_i^{\rm eq}]_2)\right]_2$,
we get
\begin{equation}\label{eq:polynomial2}
\left[ f_i^{\rm eq}\right]_2=\rho W_i\left(1+\frac{\bm{u}\cdot\bm{v}_i}{c_s^2} + \frac{(\bm{u}\cdot\bm{v}_i)^2-c_s^2(\bm{u}\cdot\bm{u})}{2c_s^4}\right).
\end{equation}
The second-order polynomial (\ref{eq:polynomial2}) generated by the equilibrium (\ref{eq:entropic_eq}) is manifestly isotropic, and with the definition of the speed of sound $c_s=1/\sqrt{3}$ (\ref{eq:sound}) it is identical to the standard lattice Boltzmann equilibrium.

\section{Moment System for D3Q27}

We consider the standard 27 velocity lattice (D3Q27). 
%
Let the natural moments be defined as 
\begin{equation}
\label{eq:natmoments}
\rho M_{pqr}= \integral{f_i v_{ix}^p v_{iy}^q v_{iz}^r},\ ~ p,q,r \in \lbrace0,1,2\rbrace
\end{equation}
where notation $\integral{...}$ is used as a shorthand for summation over all velocity indices. 
In the basis spanned by the natural moments populations can be represented as 
\begin{widetext}
\begin{align}
\label{eq:moment_rep}
\begin{split}
f_{( 0, 0, 0)}                 &= \rho              \left(1                                   - T + M_{022}+M_{202}+M_{220}-M_{222}\right) \\
f_{(\sigma, 0, 0)}             &= \tfrac{1}{6} \rho \left(3\sigma u_x   + 2 N_{xz} -   N_{yz} + T - 3\sigma  Q_{xyy} - 3 \sigma  Q_{xzz} + 3\sigma  M_{122} - 3 M_{202} - 3 M_{220} + 3 M_{222}\right) \\
f_{( 0, \lambda, 0)}           &= \tfrac{1}{6} \rho \left(3\lambda u_y  -   N_{xz} + 2 N_{yz} + T - 3\lambda Q_{xxy} - 3 \lambda Q_{yzz} + 3\lambda M_{212} - 3 M_{022} - 3 M_{220} + 3 M_{222}\right) \\
f_{( 0, 0, \delta)}            &= \tfrac{1}{6} \rho \left(3\delta u_z   -   N_{xz} -   N_{yz} + T - 3\delta  Q_{xxz} - 3 \delta  Q_{yyz} + 3\delta  M_{221} - 3 M_{022} - 3 M_{202} + 3 M_{222}\right) \\
f_{( \sigma, \lambda, 0)}      &= \tfrac{1}{4} \rho \left(\sigma \lambda \Pi_{xy} + \lambda Q_{xxy} + \sigma  Q_{xyy} + M_{220} - \sigma  M_{122} - \lambda M_{212} - \sigma  \lambda M_{112} - M_{222}\right)  \\
f_{( \sigma, 0, \delta)}       &= \tfrac{1}{4} \rho \left(\sigma \delta  \Pi_{xz} + \delta  Q_{xxz} + \sigma  Q_{xzz} + M_{202} - \sigma  M_{122} - \delta  M_{221} - \sigma  \delta  M_{121} - M_{222}\right) \\
f_{( 0, \lambda, \delta)}      &= \tfrac{1}{4} \rho \left(\lambda \delta \Pi_{yz} + \delta  Q_{yyz} + \lambda Q_{yzz} + M_{022} - \lambda M_{212} - \delta  M_{221} - \lambda \delta  M_{211} - M_{222}\right) \\
f_{( \sigma, \lambda, \delta)} &= \tfrac{1}{8} \rho \left(\sigma \lambda \delta Q_{xyz}                                         + \sigma  M_{122} + \lambda M_{212} + \delta  M_{221} + \sigma \lambda M_{112} + \sigma \delta M_{121} + \lambda\delta M_{211} + M_{222}\right).
\end{split}
\end{align}
%
%
\end{widetext}
Here we depart form the usual single index $i$ and use a somewhat unconventional 
indexing for the discrete velocities, where indices $\sigma,\lambda,\gamma\in\lbrace-1,1\rbrace$. 
Note that labelling of the velocities by a triad $(\cdot,\cdot,\cdot)$ is unambiguous as long as the first, the second, and the third entries are always associated with the $x$, the $y$, and the $z$ coordinates, respectively, in the once fixed Cartesian frame.

We chose to rename some of the natural moments as reminiscence to their physical meaning:
$$T=M_{200}+M_{020}+M_{002}$$ 
is the trace of the stress tensor at unit density, 
\begin{align*}
N_{xz}&=M_{200}-M_{002}\\
N_{yz}&=M_{020}-M_{002}
\end{align*}
are the normal stress differences at unit density and 
\begin{align*}
\Pi_{xy}&=M_{110}\\
\Pi_{xz}&=M_{101}\\
\Pi_{yz}&=M_{011}
\end{align*}
are the off-diagonal
components of the stress tensor at unit density. The third order moments lack a direct physical interpretation in the isothermal case
but are denoted as $Q_{xzz}=M_{102}$, $Q_{xxy}=M_{210}$, $Q_{yyz}=M_{021}$, $Q_{xxz}=M_{201}$, $Q_{yyz}=M_{021}$ and $Q_{xyz}=M_{111}$.

Another basis is given by the central moments of the form
\begin{equation}
\rho\widetilde{M}_{pqr}=\integral{f_{i} (v_{ix}-u_x)^p (v_{iy}-u_y)^q (v_{iz}-u_z)^r}.
\end{equation}
Using the mapping from natural to central moments ~\eqref{eq:cm1}-\eqref{eq:cm23}, which is linear in the non-conserved moments, a similar moment representation of the populations in the central moments basis can then be written~\eqref{eq:cmf000}-\eqref{eq:cmfxyz}. 

\section{KBC}

Let us then group the contribution to each population in three parts 
\begin{equation}\label{eq:ksh-representation}
f_i=k_i+s_i+h_i,
\end{equation}
where $k_i$ (= kinematic part) depends only on the locally conserved fields, $s_i$ (= shear part) depends on the (deviatoric) stress tensor 
$\bm{\Pi}_0$
%
%
and $h_i$ (= higher-order moments) is a linear combination of the remaining higher-order moments. 
Representation~\eqref{eq:ksh-representation} is easily obtained for any lattice and any moment basis. 

With the representation~\eqref{eq:ksh-representation}, a different mirror state can be sought in a one-parameter form,
\begin{equation}\label{eq:ksh-mirror}
f_i^{\rm mirr}=k_i+[2s_i^{\rm eq}-s_i]+[(1-\gamma)h_i+\gamma h_i^{\rm eq}],
\end{equation}
where $\gamma$ is a parameter which is not yet specified and the terms $s_i^{\rm eq}$ and $h_i^{\rm eq}$ denote the $s$ and $h$ part evaluated at equilibrium \eqref{eq:entropic_eq}.
When \eqref{eq:ksh-mirror} is used in \eqref{eq:Egeneral}, one arrives at nothing but a special (not the most general) MRT model. 
For { any} $\gamma$, the resulting LB model still recovers hydrodynamics with the same kinematic viscosity 
$\nu$ \eqref{eq:viscosity}. 
Note that Eq.\ \eqref{eq:Egeneral} and Eq.\ \eqref{eq:ksh-mirror} can be also rewritten as 
$$f'_i = f_i + 2 \beta (f_i^{\rm GE} - f_i),$$
with the generalized equilibrium  \cite{Asinari09,Asinari10,Asinari11} of the form,
$$f_i^{\rm GE} = f_i^{\rm eq} + (1/2) (\gamma - 2) (h_i^{\rm eq} - h_i).$$
%
For $\gamma=2$ we obtain the LBGK model while the choice of $\gamma=\tfrac{1}{\beta}$ results in the generalized family of regularized LB (RLB) models.

While the dependence of $s$ on the deviatoric stress tensor $\bm{\Pi}_0$ is mandated, one may include further moments, provided the basic symmetry properties are not violated. We will consider eight different KBC models, all of which give the correct kinematic viscosity in the hydrodynamic limit but differ in the choice of $s_i$. 


The four natural moment KBC models considered herein are characterized by 
$$s \in \left\lbrace d,\ d+t,\ d+q,\ d+t+q \right\rbrace,$$
where $d$ depends on the deviatoric stress tensor $\bm{\Pi}_0$, $t$ depends on the trace of the stress tensor $T = {\rm tr}(\bm{\Pi})$ and $q$ is a function of the third order moments $Q_{\alpha\beta\gamma}$. Table~\ref{tab:s_parts} shows a comprehensive list for $d$, $t$, $q$ and the kinematic part $k$ in the natural moments representation. Note that the model with $s=d+t$ and fixed $\gamma=\tfrac{1}{\beta}$ is equivalent to the regularized LB model in \cite{Latt2006}.

Analogously, we define four KBC variations in the central moment basis, 
$$s \in \left\lbrace \widetilde{d},\ \widetilde{d}+\widetilde{t},\ \widetilde{d}+\widetilde{q},\ \widetilde{d}+\widetilde{t}+\widetilde{q} \right\rbrace$$ 
where $\widetilde{k}$, $\widetilde{d}$, $\widetilde{t}$ and $\widetilde{q}$ correspond to the central moments representations~\eqref{eq:cmf000}-\eqref{eq:cmfxyz}.
The remaining higher order parts are trivially given as $$h=f-k-s$$ in all cases.

\begin{table}
\centering
\begin{tabular}{ccccc}
\toprule
 index& $k/\rho$ & $d/\rho$ & $t/\rho$ & $q/\rho$ \\
\colrule
$( 0, 0, 0)$                 & $1$ & $0$ & $-T$ & $0$ \\
$(\sigma, 0, 0)$             & $\sigma  u_x/2$ & $(2 N_{xz} -   N_{yz} )/6$  & $T/6$ & $-\sigma  (Q_{xyy} + Q_{xzz})/2$ \\
$( 0, \lambda, 0)$           & $\lambda u_y/2$ & $(-   N_{xz} + 2 N_{yz})/6$ & $T/6$ & $-\lambda (Q_{xxy} + Q_{yzz})/2$ \\
$( 0, 0, \delta)$            & $\delta  u_z/2$ & $(-   N_{xz} -   N_{yz})/6$ & $T/6$ & $-\delta  (Q_{xxz} + Q_{yyz})/2$ \\
$( \sigma, \lambda, 0)$      & $0$ & $\sigma  \lambda \Pi_{xy}/4$ & $0$ & $( \lambda Q_{xxy} + \sigma  Q_{xyy})/4$ \\
$( \sigma, 0, \delta)$       & $0$ & $\sigma  \delta  \Pi_{xz}/4$ & $0$ & $( \delta  Q_{xxz} + \sigma  Q_{xzz})/4$ \\
$( 0, \lambda, \delta)$      & $0$ & $\lambda \delta  \Pi_{yz}/4$ & $0$ & $( \delta  Q_{yyz} + \lambda Q_{yzz})/4$ \\
$( \sigma, \lambda, \delta)$ & $0$ & $0$ & $0$ & $\sigma \lambda \delta Q_{xyz}/8$ \\
\botrule
\end{tabular}
\caption{Moment groups in the natural basis listed for all discrete velocity directions. $k$ is the kinematic part, $d$ is a function of the moments amounting to the deviatoric stress tensor, $t$ is a function of $T = {\rm tr}(\bm{\Pi})$  associated with the fluid bulk viscosity and $q$ contains the contributions of the third order moment rank-3 tensor.}
\label{tab:s_parts}
\end{table}

The major change of perspective here is that the stabilizer $\gamma$ should not be considered as a tunable parameter. 
Rather, it has to be put under entropy control and computed  by maximizing the entropy in the post-collision state $f'$. 
This matches the physics of the problem at hand, since constrained equilibria  correspond to the maximum of the entropy (here the constraint is that the $s$ part remains fixed by the over-relaxation, $s_i^{\rm mirr}=2s_i^{\rm eq}-s_i$). 

Specifically, let $S(\gamma)$ be the entropy of the post-collision states appearing in the right hand side of \eqref{eq:Egeneral}, with the mirror state \eqref{eq:ksh-mirror}. Then we require that the stabilizer $\gamma$ corresponds to maximum of this function.
Introducing deviations, $\Delta s_i=s_i-s_i^{\rm eq}$ and $\Delta h_i=h_i-h_i^{\rm eq}$, the condition for the critical point reads:
\begin{equation}\label{eq:result1}
\sum_{i=1}^{b}\Delta h_i\ln\!\left(1+\frac{(1-\beta\gamma)\Delta h_i-(2\beta-1)\Delta s_i}{f_i^{\rm eq}}\right)=0.
\end{equation}
Equation \eqref{eq:result1} suggests that among all non-equilibrium states with the fixed mirror values $s_i^{\rm mirr}=2s_i^{\rm eq}-s_i$, we pick the one which maximizes the entropy. In contrast to MRT, the entropic stabilizer $\gamma$ is not tunable but is rather computed at each lattice site in every time step from Eq.\ \eqref{eq:result1}. 
Thus, the entropic stabilizer self-adapts to a  value given by the maximum entropy condition \eqref{eq:result1}. 


In order to clarify the properties of the solution to Eq.\ \eqref{eq:result1}, let us introduce the  entropic scalar product $\langle X{|}Y\rangle$ in the $b$-dimensional vector space,
\begin{equation}
\langle X{|}Y\rangle=\sum_{i=1}^{b}\frac{X_iY_i}{f_i^{\rm eq}},
\end{equation}
and expand in \eqref{eq:result1} to the first non-vanishing order in $\Delta s_i/f_i^{\rm eq}$ and $\Delta h_i/f_i^{\rm eq}$ to obtain
\begin{equation}\label{eq:result21}
\gamma^*= \frac{1}{\beta}-\left(2-\frac{1}{\beta}\right)\frac{\langle\Delta s{|}\Delta h\rangle}{\langle\Delta h{|}\Delta h\rangle}.
\end{equation}
The result \eqref{eq:result21} explains the mechanism of failure of the proposal $\gamma\approx 1$ at $\beta\approx 1$: Whenever vectors $\Delta s$ and $\Delta h$ are non-orthogonal (in the sense of the entropic scalar product), the deviation of $\gamma^*$ from $\gamma=1$ may become very significant. 
Indeed, in \eqref{eq:result21}, the correlation between the shear and the higher-order parts  $\sim \langle\Delta s|\Delta h\rangle$ is not  a correction to $\gamma=1$ but rather a contribution of { same order} $O(1)$.

We have found that the estimate \eqref{eq:result21} was sufficient for stabilizing all the simulations which renders the KBC an explicit and efficient method with only slightly increased computational costs compared to the standard LBGK.

The resulting collision operation is given here:
\begin{lstlisting}[mathescape]
1 compute conserved quantities $\displaystyle \rho, u_\alpha$
2 evaluate equilibrium $f_i^{\rm eq} ( \rho, u_\alpha )$
3 compute $s$ and $s^{\rm eq}$, i.e. $s=d+t+q$, $s^{\rm eq}=d^{\rm eq}+t^{\rm eq}+q^{\rm eq}$ (see table @\ref{tab:s_parts}@)
4 compute $\Delta s_i = s_i - s_i^{\rm eq}$
5 compute $\Delta h_i = h_i - h_i^{\rm eq} = f_i -f_i^{\rm eq} - \Delta s_i$
6 evaluate $\gamma$ from equation @\eqref{eq:result21}@
7 relax $f_i^\prime = f_i - \beta\left( 2 \Delta s_i + \gamma \Delta h_i \right)$
\end{lstlisting}

\begin{widetext}

\begin{table}
\centering
\begin{tabular}{l|ccccc|ccccc|ccccc}
\toprule
 & \multicolumn{5}{c|}{$t=0.25$} &  \multicolumn{5}{c|}{$t=0.5$} &  \multicolumn{5}{c}{$t=0.75$} \\
                                      & A       & B       & C       & \textbf{D}   & $p$      & A       & B       & C       & \textbf{D}   & $p$    & A       & B       & C       & \textbf{D}   & $p$ \\
\colrule
$k \cdot 10^4$                        & $8.528$ & $8.644$ & $8.663$ & $\bm{8.657}$ & $2.21$   & $6.237$ & $6.385$ & $6.411$ & $\bm{6.402}$ & $2.10$ & $3.808$ & $3.947$ & $3.965$ & $\bm{3.954}$ & $1.87$ \\ 
$\Omega N^2$                          & $1.472$ & $1.735$ & $1.784$ & $\bm{1.784}$ & $2.67$   & $1.919$ & $2.786$ & $3.173$ & $\bm{3.216}$ & $2.46$ & $1.551$ & $2.130$ & $2.262$ & $\bm{2.301}$ & $2.13$ \\ 
$\epsilon N \cdot 10^5$               & $2.459$ & $2.892$ & $2.974$ & $\bm{2.974}$ & $2.65$   & $3.247$ & $4.646$ & $5.288$ & $\bm{5.360}$ & $2.44$ & $2.646$ & $3.553$ & $3.770$ & $\bm{3.835}$ & $2.10$ \\ 
\colrule
$S_3 \cdot 10^2$                      & $10.73$ & $7.361$ & $9.071$ & $\bm{10.34}$ & $1.23^*$ & $28.91$ & $38.86$ & $47.78$ & $\bm{48.97}$ & $2.04$ & $29.37$ & $33.81$ & $39.71$ & $\bm{39.99}$ & $2.62$ \\ 
$S_4$                                 & $5.075$ & $6.188$ & $6.176$ & $\bm{6.149}$ & $2.66$   & $3.993$ & $5.218$ & $6.063$ & $\bm{6.159}$ & $2.25$ & $4.140$ & $4.319$ & $4.840$ & $\bm{4.842}$ & $4.23$ \\ 
$S_5$                                 & $2.902$ & $2.157$ & $2.472$ & $\bm{2.891}$ & $0.81^*$ & $3.210$ & $6.084$ & $9.044$ & $\bm{9.690}$ & $1.66$ & $4.337$ & $4.313$ & $6.074$ & $\bm{5.978}$ & $2.05$ \\ 
$S_6$                                 & $52.50$ & $83.87$ & $81.18$ & $\bm{79.34}$ & $1.93$   & $34.07$ & $66.86$ & $96.83$ & $\bm{100.1}$ & $2.17$ & $38.64$ & $36.27$ & $51.73$ & $\bm{51.55}$ & $3.08$ \\ 
\colrule
$l_0/N \cdot 10^1$                    & $10.13$ & $8.788$ & $8.573$ & $\bm{8.564}$ & $3.72$   & $4.796$ & $3.473$ & $3.070$ & $\bm{3.022}$ & $2.60$ & $2.809$ & $2.207$ & $2.094$ & $\bm{2.050}$ & $2.05$ \\ 
$u_0 \cdot 10^2$                      & $2.920$ & $2.940$ & $2.943$ & $\bm{2.942}$ & $2.23$   & $2.497$ & $2.527$ & $2.532$ & $\bm{2.530}$ & $2.02$ & $1.951$ & $1.987$ & $1.991$ & $\bm{1.988}$ & $1.81$ \\ 
$\tau_0/N$                            & $34.68$ & $29.89$ & $29.13$ & $\bm{29.11}$ & $4.06$   & $19.21$ & $13.74$ & $12.12$ & $\bm{11.94}$ & $2.67$ & $14.39$ & $11.11$ & $10.52$ & $\bm{10.31}$ & $2.14$ \\ 
${\rm Re}_0$                          & $3549$  & $3101$  & $3028$  & $\bm{3024}$  & $3.52$   & $1437$  & $1053$  & $932.7$ & $\bm{917.6}$ & $2.55$ & $657.8$ & $526.2$ & $500.3$ & $\bm{489.2}$ & $1.96$ \\ 
\colrule
$\lambda/N \cdot 10^2$                & $5.376$ & $4.991$ & $4.927$ & $\bm{4.925}$ & $3.91$   & $4.001$ & $3.384$ & $3.179$ & $\bm{3.155}$ & $2.57$ & $3.463$ & $3.043$ & $2.960$ & $\bm{2.931}$ & $2.10$ \\ 
$u_\lambda \cdot 10^2$                & $2.384$ & $2.401$ & $2.403$ & $\bm{2.402}$ & $2.08$   & $2.039$ & $2.063$ & $2.067$ & $\bm{2.066}$ & $2.38$ & $1.593$ & $1.622$ & $1.626$ & $\bm{1.624}$ & $1.98$ \\ 
$\tau_\lambda/N$                      & $2.254$ & $2.079$ & $2.050$ & $\bm{2.050}$ & $2.81$   & $1.962$ & $1.640$ & $1.537$ & $\bm{1.527}$ & $2.72$ & $2.174$ & $1.876$ & $1.821$ & $\bm{1.805}$ & $2.26$ \\ 
${\rm Re}_\lambda$                    & $153.8$ & $143.8$ & $142.1$ & $\bm{142.0}$ & $3.44$   & $97.89$ & $83.79$ & $78.86$ & $\bm{78.21}$ & $2.46$ & $66.22$ & $59.23$ & $57.75$ & $\bm{57.11}$ & $1.92$ \\ 
\colrule
$\eta/N \cdot 10^3$                   & $2.202$ & $2.115$ & $2.100$ & $\bm{2.100}$ & $2.77$   & $2.055$ & $1.879$ & $1.819$ & $\bm{1.813}$ & $2.67$ & $2.163$ & $2.009$ & $1.979$ & $\bm{1.971}$ & $2.29$ \\ 
$u_\eta \cdot 10^3$                   & $3.784$ & $3.940$ & $3.968$ & $\bm{3.968}$ & $2.72$   & $4.056$ & $4.436$ & $4.582$ & $\bm{4.597}$ & $2.59$ & $3.853$ & $4.148$ & $4.210$ & $\bm{4.228}$ & $2.19$ \\ 
$\tau_\eta/N \cdot 10^1$              & $5.821$ & $5.368$ & $5.293$ & $\bm{5.293}$ & $2.82$   & $5.066$ & $4.235$ & $3.970$ & $\bm{3.943}$ & $2.69$ & $5.612$ & $4.843$ & $4.701$ & $\bm{4.662}$ & $2.30$ \\ 
\botrule
\end{tabular}
\caption{Comparison of LBGK and KBC ($s=d+t+q$) for statistical quantities in Kida vortex flow at ${\rm Re}=\tfrac{U_0 N}{\nu}=6000$ and 
$t=\tfrac{N}{U_0}=0.25,0.5,0.75$. Resolutions $N=100, 200, 400, 600$ for ${\rm KBC}$ runs A, B, C and resolved ${\rm LBGK}$ run D, respectively. Convergence rate $p$ of error w.r.t. LBGK solution estimated from polynomial fit ($^*$ indicates exclusion of lowest resolution). \\
All gradient-based quantities are computed with eighth-order of accuracy central difference stencils.\\
Turbulence characteristics: length scale $l_0=k^{3/2}/\epsilon$, velocity scale $u_0=k^{1/2}$, time scale $\tau_0=l_0/u_0$, and Reynolds number ${\rm Re}_0=l_0 u_0/\nu$.\\
Taylor characteristics: Taylor micro scale $\lambda=\left(15\nu u^{\prime\,2}/\epsilon \right)^{1/2}$,  velocity scale $u_\lambda=u^\prime=(2k/3)^{1/2}$ (r.m.s. turbulence intensity),
time scale $\tau_\lambda = \lambda /u_\lambda$ and Reynolds number ${\rm Re}_\lambda=\lambda u_\lambda/\nu$.\\
Kolmogorov characteristics: length scale $\eta=(\nu^3/\epsilon)^{1/4}$, velocity scale $u_\eta=\left(\nu \epsilon\right)^{1/4}$ and time scale $\tau_\eta = \left( \nu/\epsilon\right)^{1/2}$.
}
\label{tab:statistics}
\end{table}

\end{widetext}

\section{Numerical Simulations}

\begin{figure}
\centering
\includegraphics[width=0.48\textwidth]{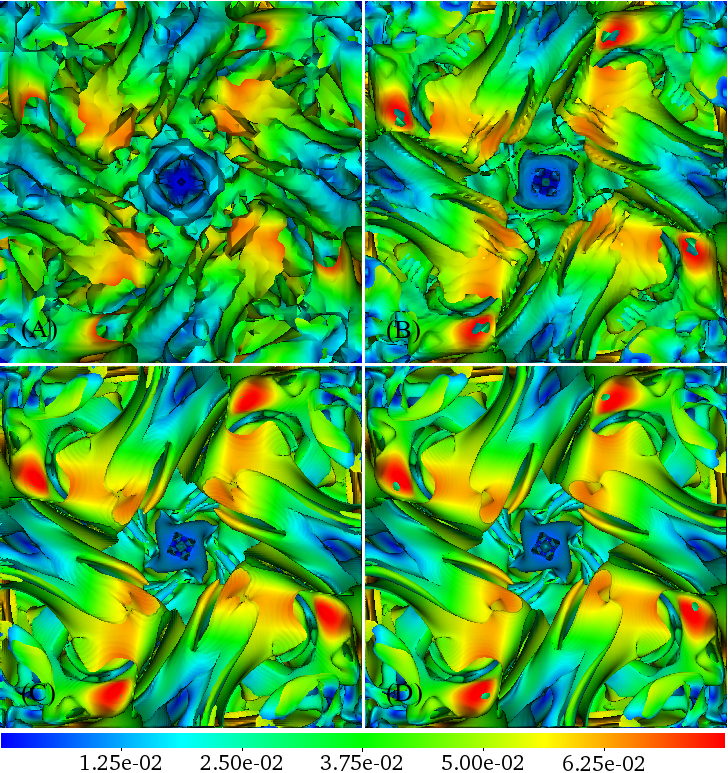}
\caption{Iso-surface of vorticity component $\omega_z = 0$ at time $t=0.5$ colored with velocity magnitude rendered at $z=0$, $x,y \in [0,\pi]$ plane. 
Runs A ($N=100$), B ($N=200$), C ($N=400$) and D ($N=600$, reference solution).}
\label{fig:vorticity}
\end{figure}

The Kida vortex flow is a well-studied benchmark flow which evolves from a simple deterministic and symmetric initial condition to a state which resembles a fully developed turbulent flow, which features a corresponding energy cascade. 
The initial conditions for the flow field are given by
\begin{equation}
\label{eq:kida}
\begin{split}
u_x(x,y,z) &= U_0 \sin x ( \cos 3 y \cos z - \cos y \cos 3 z ) \\
u_y(x,y,z) &= U_0 \sin y ( \cos 3 z \cos x - \cos z \cos 3 x ) \\
u_z(x,y,z) &= U_0 \sin z ( \cos 3 x \cos y - \cos x \cos 3 y )
\end{split}
\end{equation}
where $x,y,z \in [0,2\pi]$ and periodic boundary conditions are imposed in all directions.
The Reynolds number is defined as ${\rm Re} = U_0 N / \nu$ where N is the domain size.
Initial conditions for the density (and pressure $p=\rho c_s^2$) and higher order moments are obtained by solving the convection-diffusion equation $\frac{\partial \rho}{\partial t} + \bm{\nabla} \cdot (\rho \bm{u}_0) = D \Laplace \rho$ on the same grid beforehand until steady state is reached in a similar process as described by \cite{mei2006consistent}.


The Kida vortex flow has been analyzed extensively using DNS \cite{kida1985three,kida1987kolmogorov,keating2007entropic,Chikatamarla2010}. The evolution of enstrophy shows a steep increase in the early stage of the simulation and reaches a maximum value before it decays. For the convergence study we investigate data collected from time points around the peak of enstrophy which indicates the existence of large gradients which are often numerically challenging.
A simulation was considered stable if it run until the mean enstrophy 
$$\Omega = \tfrac{1}{2} \left\langle \left\Vert \bm{\nabla} \times \bm{u}^\prime \right\Vert^2 \right\rangle$$ 
was sufficiently decayed ($\Omega/\Omega_0 < 5 \% $) and 
$$u_\alpha ^\prime = u_\alpha-\left\langle u_\alpha \right\rangle.$$

In order to assess the stability region, the domain size $N=100$ and initial velocity $U_0=0.05$ were fixed and the Reynolds number $\rm Re$ was increased in steps of $500$.
While LBGK seized to yield sensible values at $Re \gtrsim 5000$, ELBM was always stable (tested up to ${\rm Re = 10^7}$). Likewise, all KBC models were always stable, independently of the moment basis or the choice of $s$. The only RLB models, however, which achieved the same stability are using $s=d$ and $s=\widetilde{d}$, respectively. All other six RLB models, among them the standard model $s=d+t$, were less stable than LBGK.

Accuracy of KBC scheme is studied in detail using Kida vortex flow at ${\rm Re} = 6000$. 
KBC model $s=d+t+q$ is used exemplarily in all further simulations and compared to LBGK simulation at $N=600$ (run D) 
where the flow is considered to be resolved as indicated by the Kolmogorov length scale 
$\eta=\left(\nu^3/\epsilon\right)^{1/4} \approx 1.2$ lattice units where 
$$\epsilon=\tfrac{1}{2}\nu\left\langle \left( \tfrac{\partial u_\alpha^\prime}{\partial x_\beta} + \tfrac{\partial u_\beta^\prime}{\partial x_\alpha} \right) \left( \tfrac{\partial u_\alpha^\prime}{\partial x_\beta} + \tfrac{\partial u_\beta^\prime}{\partial x_\alpha} \right) \right\rangle$$
is the rate of turbulence kinetic energy dissipation.
Resolutions $N=100,200,400$ are considered in the following (runs A, B and C, respectively). 
Convergence towards resolved LBGK simulation is reported in table~\ref{tab:statistics} for various statistical quantities.  
Unless stated otherwise all quantities are given in lattice units.

Figure~\ref{fig:vorticity} shows a comparison of the vortex structures for the four simulations roughly at the point of maximum enstrophy. The vorticity configuration is thus affected by the under-representation of large gradients in the coarse resolutions (see next section). Nevertheless, the large vortex structures are akin at all resolutions. The largest KBC simulation (run C) is hardly discriminable form the reference LBGK simulaion (run D).

\begin{figure}[h]
\centering
\includegraphics[width=0.48\textwidth]{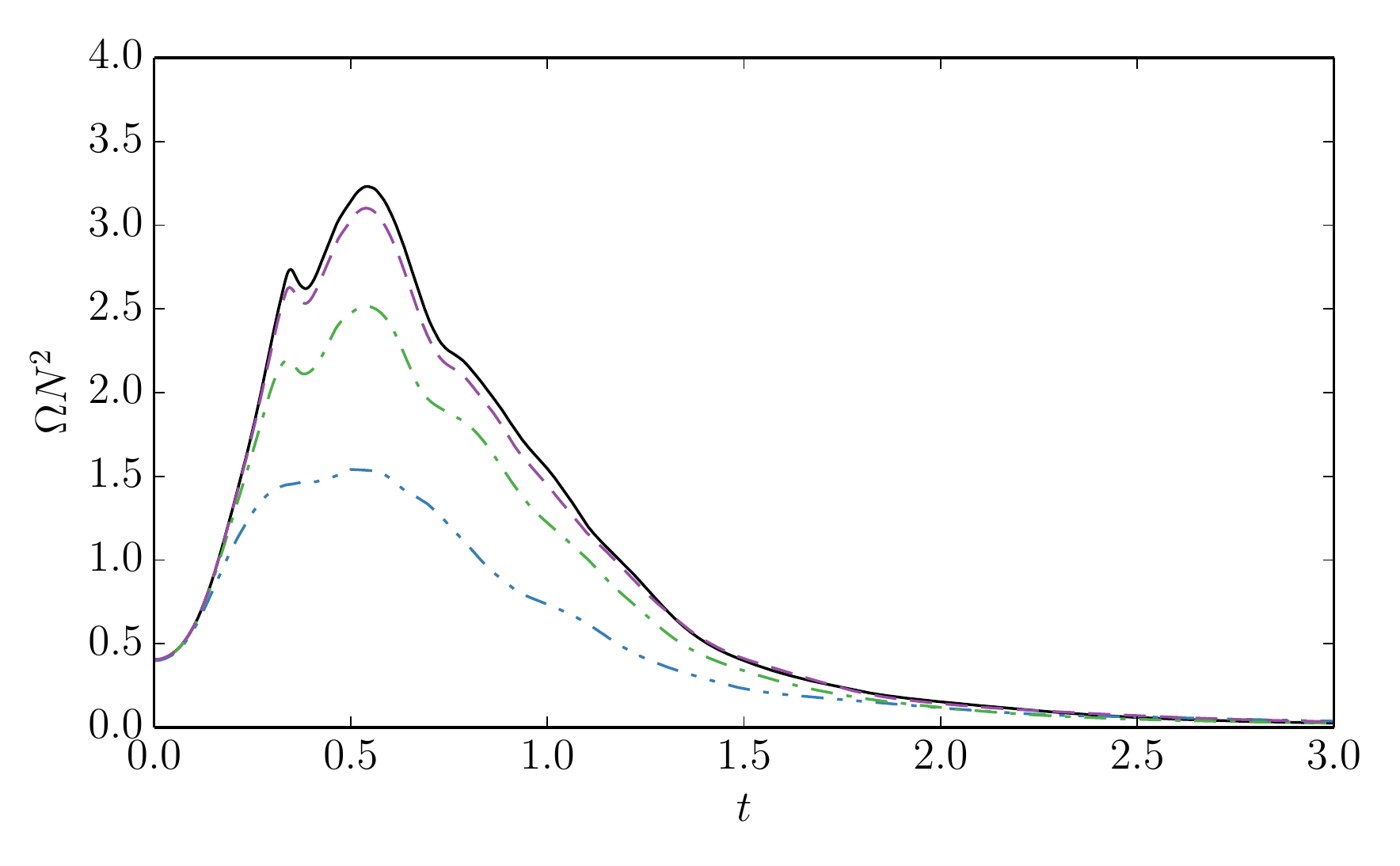}
\caption{Mean enstrophy evolution with time for simulations A (\aclr{$-\cdot \cdot$}), B (\bclr{$-\cdot$}), C (\cclr{$- -$}) and D (\fclr{$-$}). Gradients evaluated with second-order of accuracy.}
\label{fig:enstrophy}
\end{figure}

\begin{figure}[h]
\centering
\includegraphics[width=0.48\textwidth]{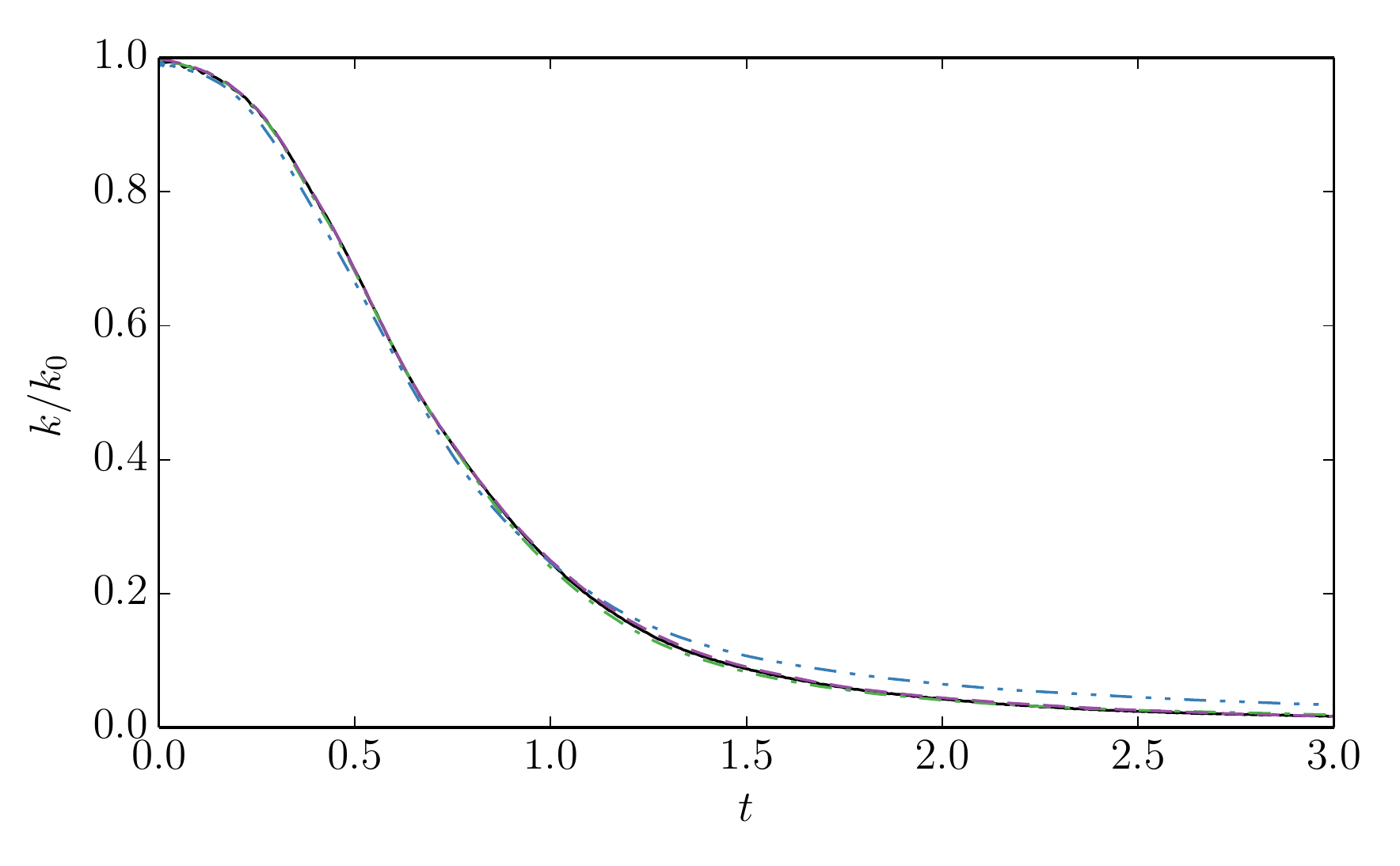}
\caption{Kinetic energy evolution with time for simulations A (\aclr{$-\cdot \cdot$}), B (\bclr{$-\cdot$}), C (\cclr{$- -$}) and D (\fclr{$-$}).}
\label{fig:energy}
\end{figure}

\begin{figure}[h]
\centering
\includegraphics[width=0.48\textwidth]{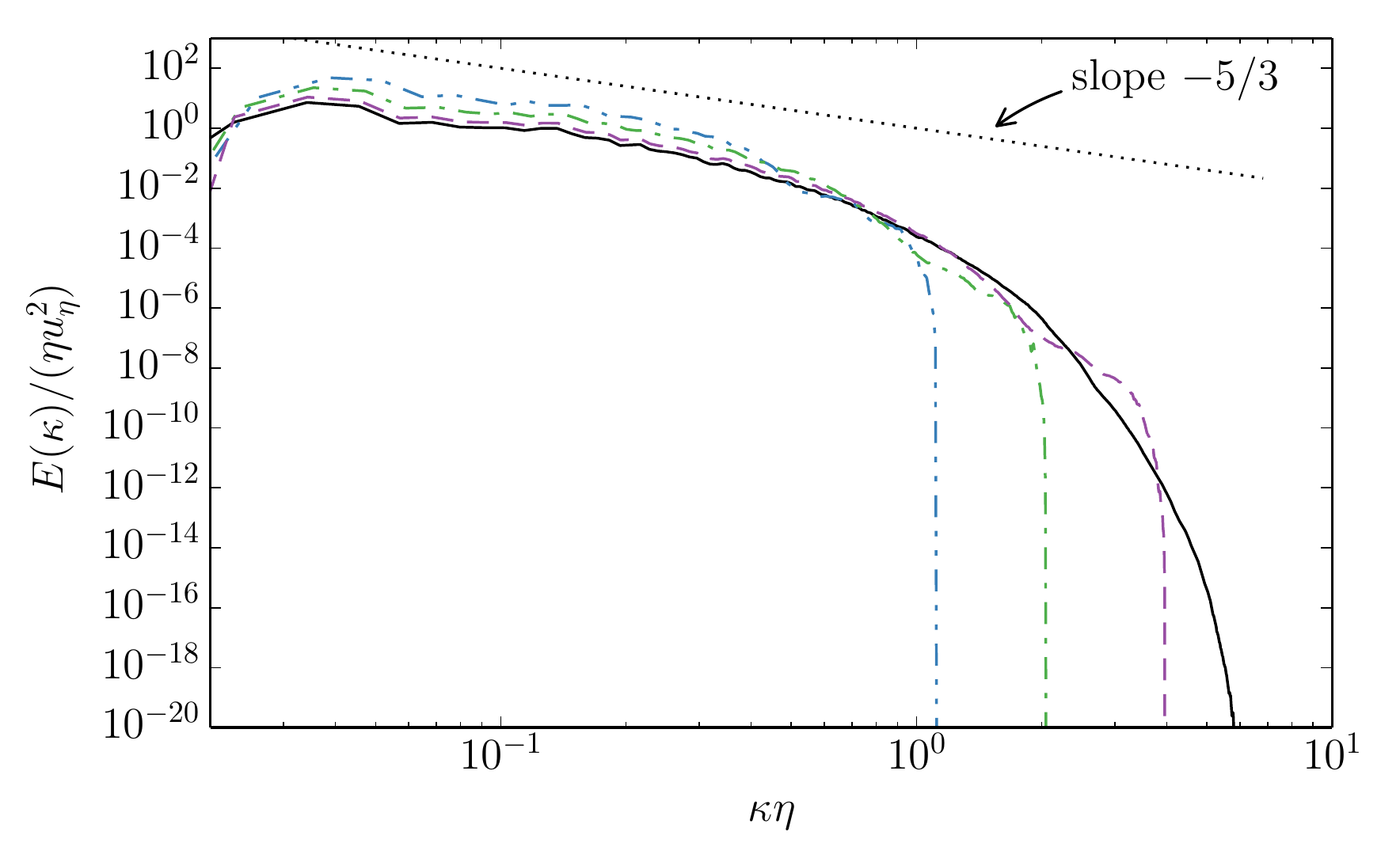}
\caption{Kinetic energy density spectrum at $t=0.5$ for simulations A (\aclr{$-\cdot \cdot$}), B (\bclr{$-\cdot$}), C (\cclr{$- -$}) and D (\fclr{$-$}). The dotted line with slope $-5/3$ shows the Kolmogorov scaling.}
\label{fig:energy-spectrum}
\end{figure}

\begin{figure}[h]
\centering
\includegraphics[width=0.48\textwidth]{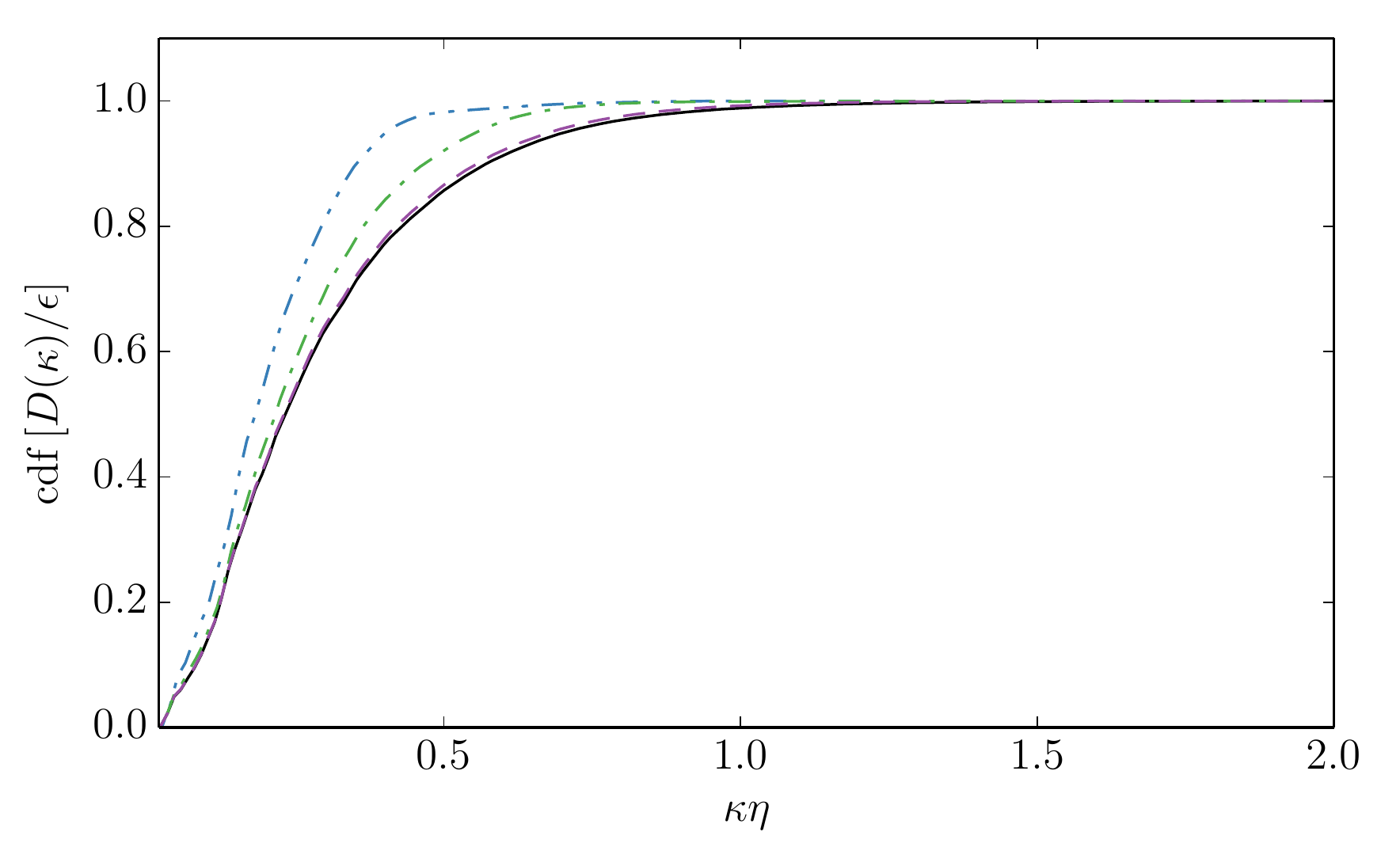}
\caption{Cumulative distribution function of the energy-dissipation rate density $D(\kappa)$ at time $t=0.5$ for simulations A (\aclr{$-\cdot \cdot$}), B (\bclr{$-\cdot$}), C (\cclr{$- -$}) and D (\fclr{$-$}).}
\label{fig:dissipation_cdf}
\end{figure}

\subsection{One-Point Statistics}

Enstrophy $\Omega$ and turbulence kinetic energy 
$$k = \tfrac{1}{2}\left\langle u_\alpha ^\prime u_\alpha ^\prime \right\rangle$$
 are important global quantities characterizing the flow and its history. Figures~\ref{fig:enstrophy} and~\ref{fig:energy}, respectively, depict the evolution of both quantities with time $t=N/U_0$. It is apparent that in the under-resolved KBC simulations $N=100, 200$ the enstrophy peak values are not well represented. However, for coarse resolutions this is expected. The kinetic energy on the other hand decays quite similarly for all simulations. Table~\ref{tab:statistics} reports the numbers at three selected time instances.
During simulation, gradients are evaluated using second-order finite differences, which are solely used for reporting the enstrophy evolution in figure~\ref{fig:enstrophy}. All quantities based on gradients in table~\ref{tab:statistics}, however, are computed with a nine point central difference stencil of eighth-order accuracy, which also explains the discrepancies in enstrophy between figure~\ref{fig:enstrophy} and table~\ref{tab:statistics} for the lowest resolution $N=100$.

While the energy seems to be dissipated similarly with time it is important to study the kinetic energy $k$ and dissipation rate $\epsilon$ thereof across flow scales in order to decide whether low-order statistics of turbulent flows yield sensible values in coarse resolution simulations despite the under-representation of high gradients. The instrument at hand is the spectral representation of the kinetic energy distribution $E(\kappa)$ where $\kappa=\Vert\bm{\kappa}\Vert$ is the modulus of the wave number vector. Figure~\ref{fig:energy-spectrum} shows the non-dimensional energy density distribution normalized with kinetic energy $$k=\int_0^\infty E(\kappa)\ d\kappa.$$ 
According to \cite{K41a,K41b} the energy scales as $E\sim \kappa^{-5/3}$ in the inertial sub-range. The studied Kida flow here does not exhibit large enough Reynolds numbers to see an extended inertial range. However, it is apparent that the energy scales similarly across resolutions and a sharp cut-off is visible at the smallest scales. This indicates that the KBC model is capable of producing the expected energy distribution throughout the scales without an explicit turbulence model. A case for higher Reynolds number shall be examined below.

The cumulative distribution function of the energy-dissipation rate density $$D(\kappa)=2\nu\kappa^2E(\kappa)$$ where $$\epsilon=\int_0^\infty D(\kappa)\ d\kappa$$ illustrates the scales of eddies responsible for the dissipative process, see figure~\ref{fig:dissipation_cdf}. The under-resolved simulations employ expectedly larger eddies for the bulk of the dissipation (see also table~\ref{tab:statistics} where $\epsilon$ is reported).

The longitudinal skewness factor $$S_{11}^n =(-1)^n \left\langle \left(\tfrac{\partial u_x^\prime}{\partial x}\right)^n\right\rangle \left\langle \left(\tfrac{\partial u_x^\prime}{\partial x}\right)^2\right\rangle^{-n/2}$$ is another global statistical quantity in real space which we report in table~\ref{tab:statistics}. In agreement with figure~\ref{fig:enstrophy} we find that the outcome of the lowest resolution $N=100$ is rather inconsistent with the trend observed in the other simulations, however, agrees well with the resolved case. The lower convergence rate for the odd-order skewness factors may be caused by the inherent lack of isotropy in the third-order moments. However, further studies are needed to draw a concise conclusion.

The remainder of table~\ref{tab:statistics} is a compilation of the turbulence, Taylor and Kolmogorov flow scales. Here and with the vast majority of the reported quantities we observe a grid convergence rate of $\approx 2$ as expected in the context of LB simulations.

\begin{figure}[t]
\centering
\includegraphics[width=0.48\textwidth]{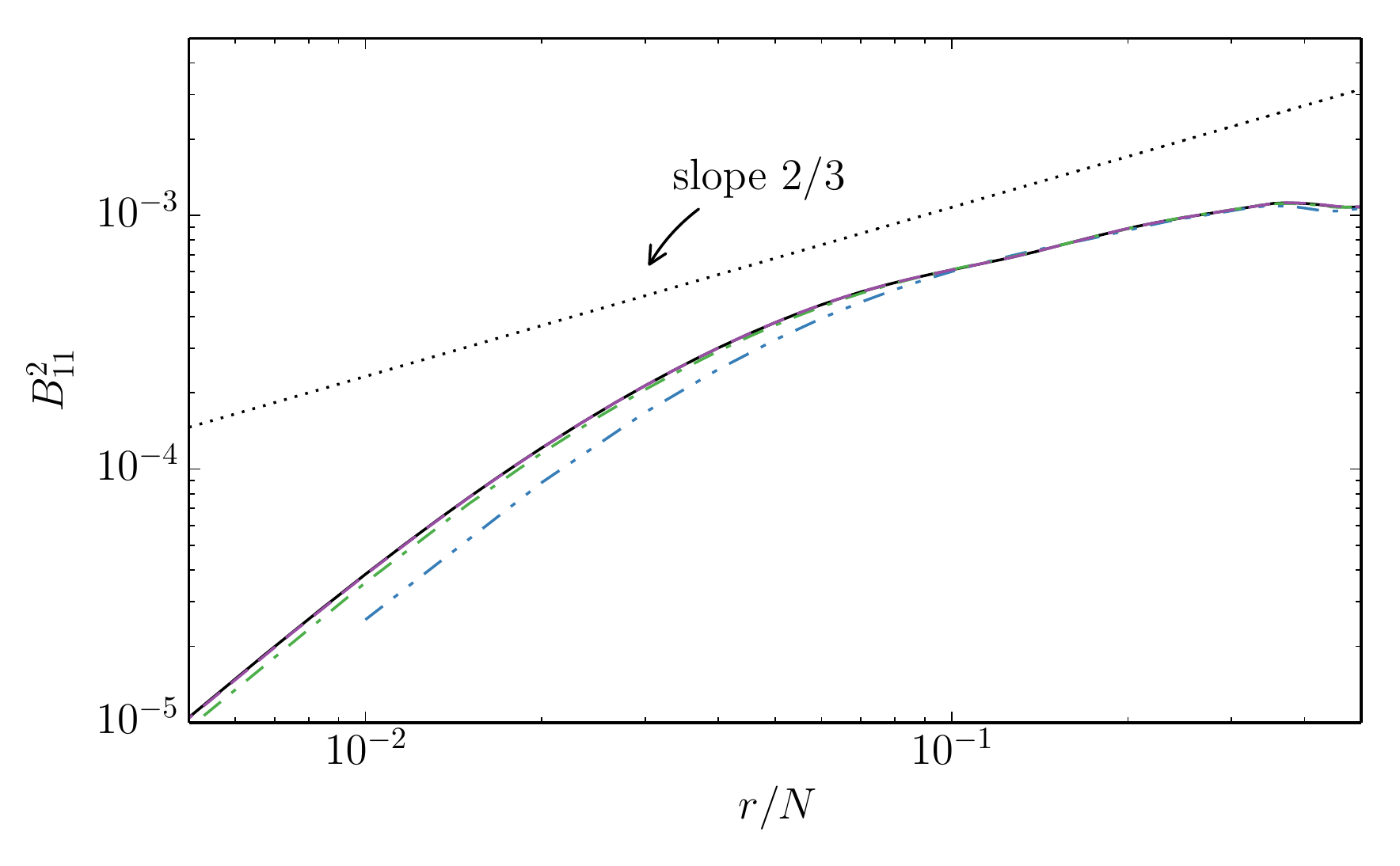}
\caption{Second order longitudinal structure function at $t=0.5$ for simulations A (\aclr{$-\cdot \cdot$}), B (\bclr{$-\cdot$}), C (\cclr{$- -$}) and D (\fclr{$-$}). The dotted line indicates the theoretical scaling.}
\label{fig:structure}
\end{figure}

\begin{figure}[h]
\centering
\includegraphics[width=0.48\textwidth]{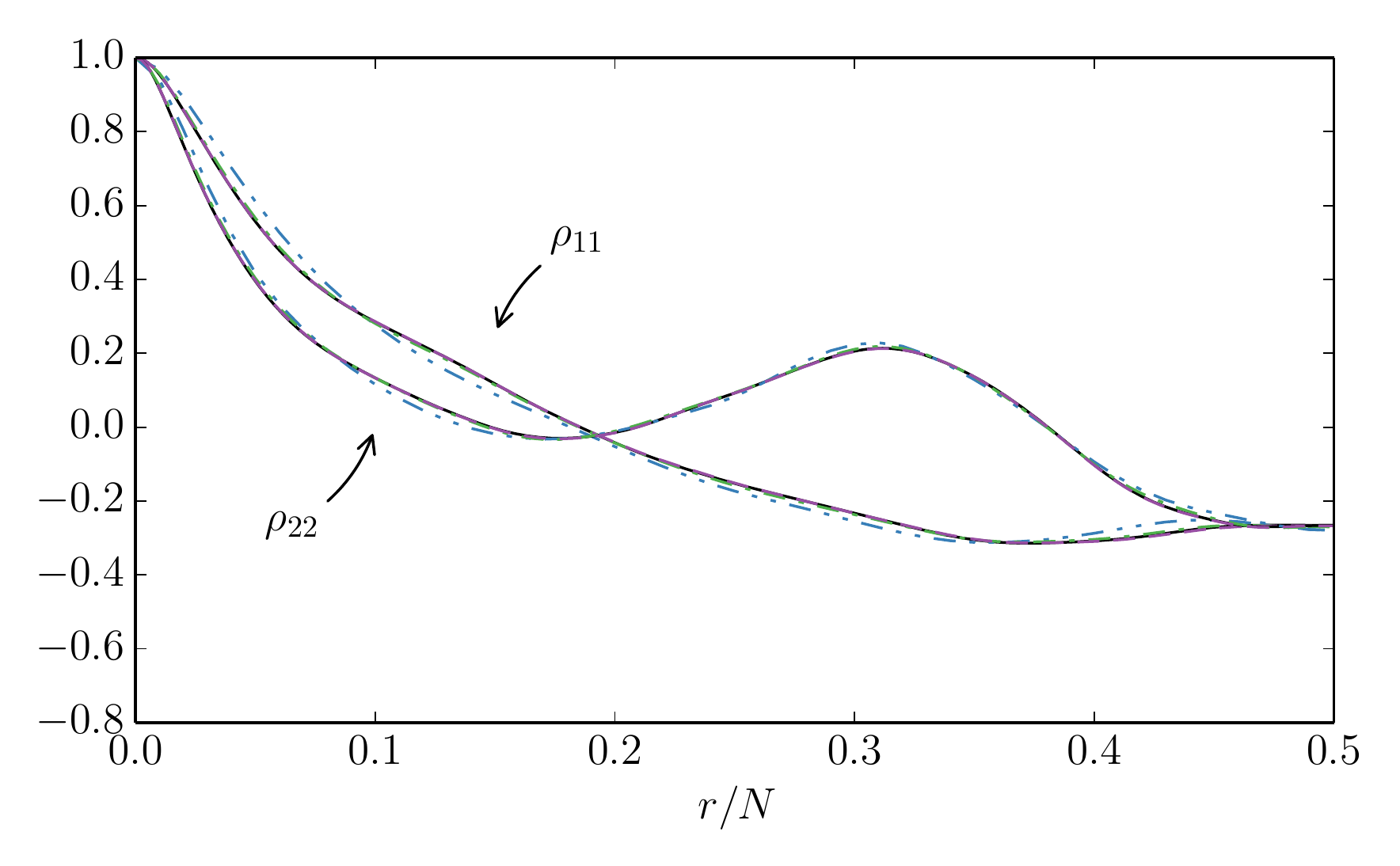}
\caption{Longitudinal and transversal velocity correlation functions at $t=0.5$ for simulations A (\aclr{$-\cdot \cdot$}), B (\bclr{$-\cdot$}), C (\cclr{$- -$}) and D (\fclr{$-$}).}
\label{fig:correlation}
\end{figure}

\subsection{Two-Point Statistics}

The longitudinal structure function of order $n$ defined as $$B_{11}^n =(-1)^n \left\langle \left(\tfrac{\partial u_x^\prime}{\partial x}\right)^n\right\rangle \left\langle \left(\tfrac{\partial u_x^\prime}{\partial x}\right)^2\right\rangle^{-n/2}$$ exhibit linear scaling on logarithmic plots \cite{K41a,K41b}. In particular, the second order structure function scales with $B_{11}^2\sim r^{2/3}$. Figure~\ref{fig:structure} depicts the results with the theoretical scaling. Due to the relatively low Reynolds number we may not identify an extended inertial range but we note that the simulations agree well with the reference over the entire range of $r$.

Another real-space two-point statistical quantity that can be used to assess different numerical techniques is the correlation of the velocity field. Here the longitudinal and transversal correlation functions are defined as 
\begin{align*}
\rho_{11}^n(r) &= \frac{\left \langle u_x^\prime(x,y,z) u_x^\prime(x+r,y,z) \right \rangle}{\left \langle u_x^\prime(x,y,z) u_x^\prime(x,y,z) \right \rangle} \\
\rho_{22}^n(r) &= \frac{\left \langle u_y^\prime(x,y,z) u_y^\prime(x+r,y,z) \right \rangle}{\left \langle u_y^\prime(x,y,z) u_y^\prime(x,y,z) \right \rangle}.
\end{align*}
A comparison at time $t=0.5$ is given in figure~\ref{fig:correlation}. All simulations with $N\geq 200$ show excellent agreement with the reference solution. At the maximum distance $r=0.5$ the velocity components are still correlated which is associated with the low Reynolds number (see figure~\ref{fig:correlation_large} for comparison).

\begin{figure}[h]
\centering
\includegraphics[width=0.48\textwidth]{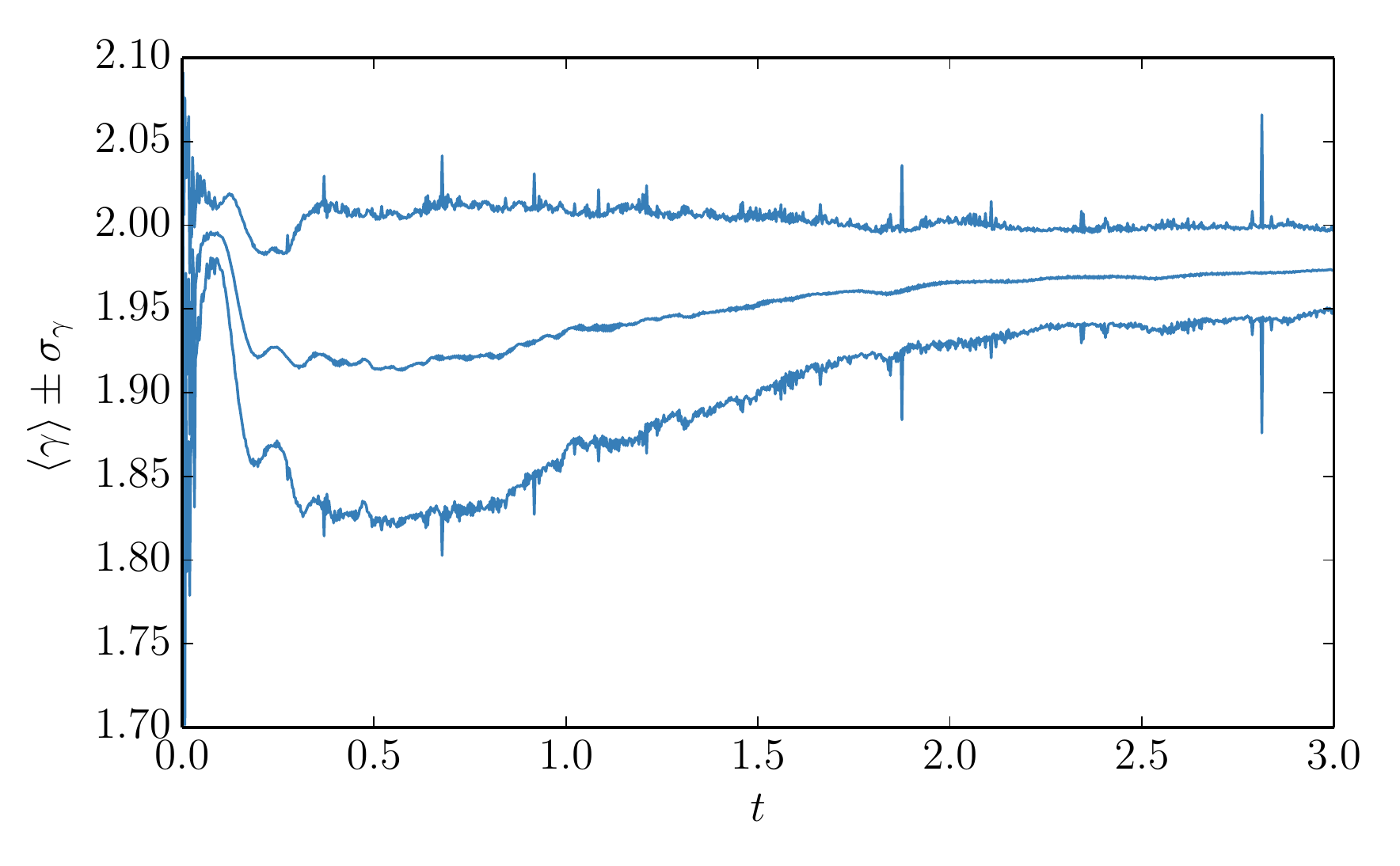}
\caption{Evolution of the mean entropic stabilizer $\langle\gamma\rangle$ and its standard deviation with time for simulation A.}
\label{fig:gamma}
\end{figure}

\subsection{Stabilizer $\gamma$}

The importance of the self-adjusting stabilizer $\gamma$ becomes clear when considering the evolution thereof with time. Figure~\ref{fig:gamma} shows the mean stabilizer $\langle\gamma\rangle$ at $N=100$ with its standard deviation. It is apparent that $\gamma$ is far form being constant. In fact, its evolution is closely correlated with the flow state. In regions of high turbulence intensity it is distinctly different in the mean and shows larger fluctuations. In \cite{KBC} we also report the distribution of $\gamma$ in space which again reveals the close relation of $\gamma$ to the flow. It is therefore conjectured, and supported by our simulations and comparisons to literature, that any MRT using $\gamma=const$ may not reach numerical stability to the same extent.

\subsection{Large Reynolds Numbers}

\begin{figure}
\centering
\includegraphics[width=0.48\textwidth]{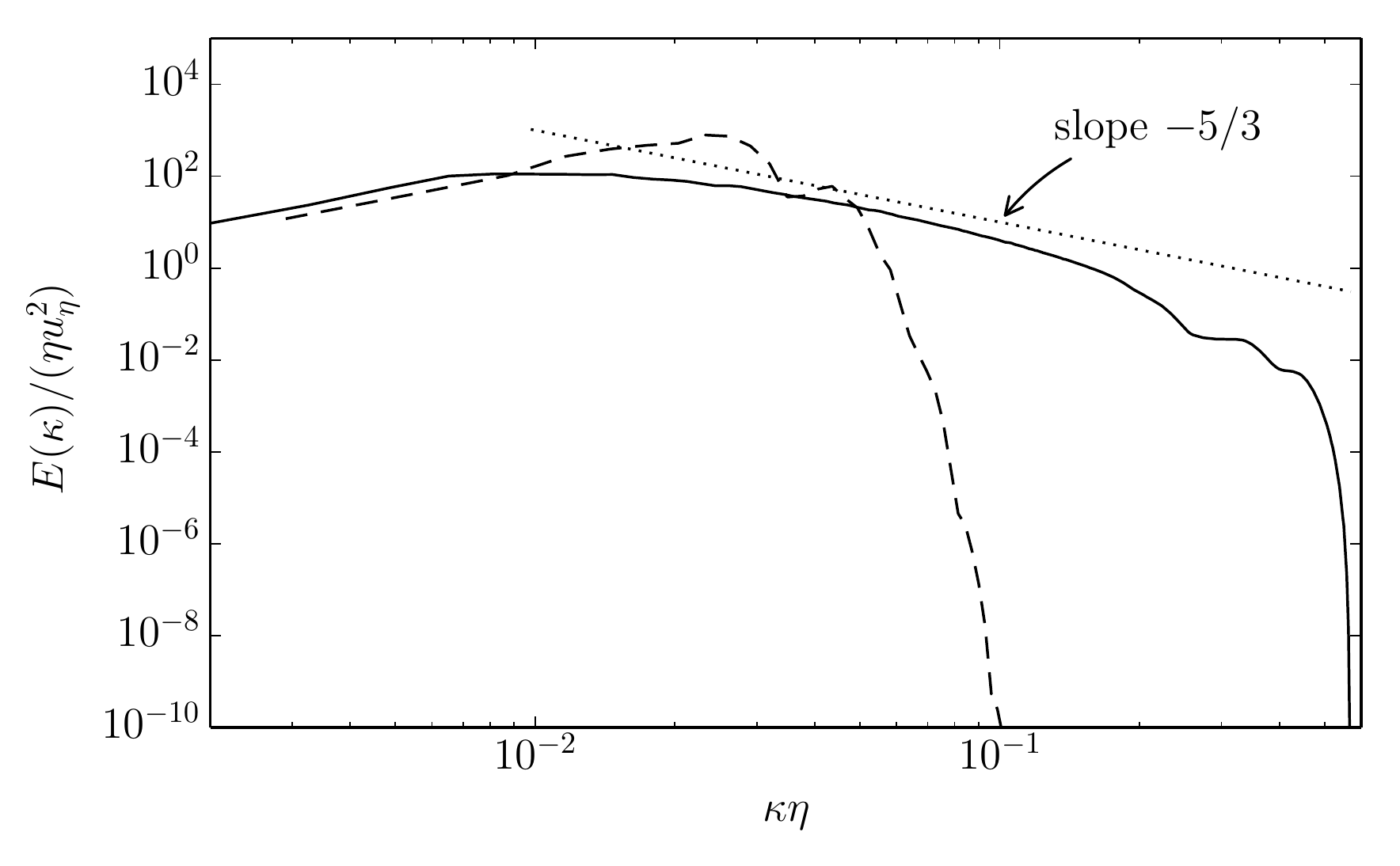}
\caption{Turbulence kinetic energy spectrum at ${\rm Re}_\lambda = 580$, $t=0.75$ (solid) and initial spectrum at ${\rm Re}_\lambda = 2000$ (dashed). The dotted line indicates the theoretical scaling.}
\label{fig:spectrum_large}
\end{figure}

\begin{figure}
\centering
\includegraphics[width=0.48\textwidth]{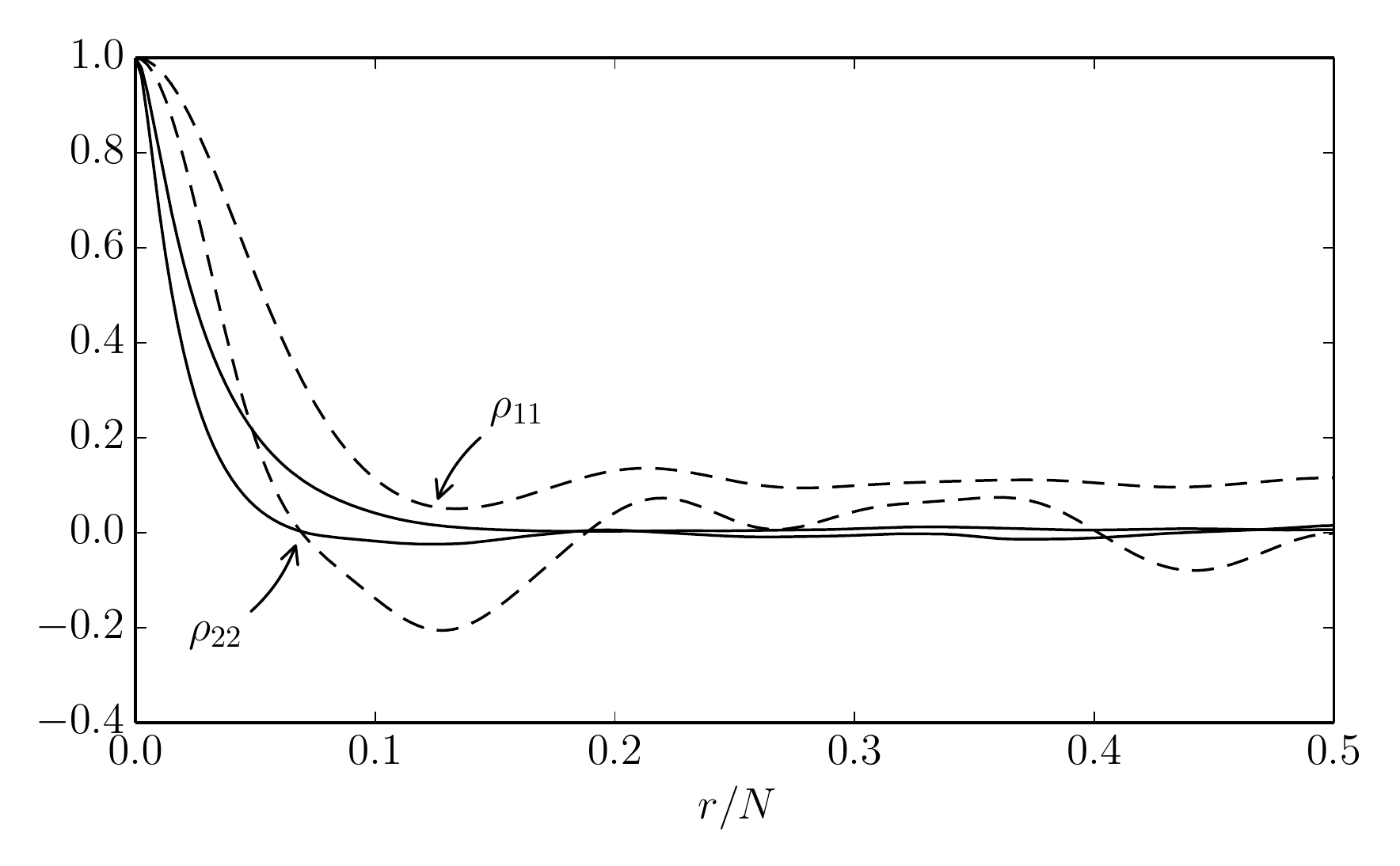}
\caption{Longitudinal and transversal velocity correlation functions at ${\rm Re}_\lambda = 580$, $t=0.75$ (solid) and initial values at ${\rm Re}_\lambda = 2000$ (dashed).}
\label{fig:correlation_large}
\end{figure}

The second numerical example considered in this paper is the simulation of a highly turbulent flow starting from a random initial condition and decaying with time. The periodic and cubic domain with box-length $N=400$ was initialized with a flow field generated from a prescribed narrow-banded initial energy spectrum peaked at grid wave-number $\tfrac{\kappa_0 N}{2\pi}\approx 8$,
%
\begin{equation}
\label{eq:E0}
\begin{split}
E_0 & =400\left(\tfrac{2}{3}\right)^{1/4} \left(2\nu {u_0^\prime}^3/\pi\right)^{1/2}\, b^2 \kappa^4 \exp{\left[-b \kappa^2\right]}, \\
b &= 20000 \left(\tfrac{2}{3}\right)^{1/2} \nu/u_0^\prime,
\end{split}
\end{equation}
where $\nu=8.164970\cdot 10^{-5}$ and $u_0^\prime =0.01$. The initial values of the density (and pressure) and the higher order moments were generated with the identical procedure as described earlier for the Kida flow.

The main objective is to test the KBC scheme for an under-resolved simulation (Kolmogorov length $\eta\approx 10$ lattice units) at large Reynolds numbers (${\rm Re}_\lambda \approx 600$). 
In particular, we ask whether the scheme is stable for a random and highly turbulent flow in absence of a deterministic and highly symmetric initial condition, whether low-order statistics are well represented and physical dissipation (i.e. scaling laws) is modelled correctly. 
By means of this simulation we examine the general question of the performance for large Reynolds numbers in an under-resolved simulation from yet another point of view. While a resolved simulation was not attempted, we compare our results to the classical scaling laws.

Figure~\ref{fig:spectrum_large} shows the turbulence kinetic energy spectrum at $t=N/u_0^\prime=0.75$. The inertial range is extended and the scaling is more apparent than in the less turbulent simulations above. The sharp cut-off at the smallest scales is still maintained despite the coarse resolution. As before, numerical stability is naturally guaranteed to very high Reynolds numbers. While the initial spectrum is narrow and steep, it flattens during the course of energy decay and exhibits the Kolmogorov scaling in the inertial sub-range roughly at peak of mean enstrophy. Figure~\ref{fig:correlation_large} shows the velocity correlations where 
the contributions to the correlations are vanishing for $r/N>0.2$ at $t=0.75$. Hence, the velocity field is largely uncorrelated as one would expect from isotropic homogeneous turbulent flows. While these results are far from a comprehensive study, they contribute to the overall assessment that the KBC scheme might perform well even in the case of severe under-resolution. A more comprehensive investigation 
shall be conducted in a further study.


\section{Conclusions}

We present three-dimensional realizations of the KBC model for the D3Q27 lattice. 
We review the details of the entropic stabilization and describe eight variations of KBC. Stability and accuracy is studied in detail for homogeneous isotropic turbulence. A detailed 
comparison with LBGK is carried out at various grid resolutions. Second order rate of convergence is numerically confirmed in the vast majority of the statistical quantities of interest.
It must be stressed that the entropic KBC models are stable (in contrast to LBGK and RLB) for all the considered cases here;
despite of under-resolution and high Reynolds numbers. 

The KBC models were shown to capture the expected scaling law for energy spectra 
in the case of high Reynolds numbers. Low order statistics such as averages of kinetic energy, enstrophy and rate of dissipation as well as the spectral densities for energy and rate of dissipation agree well with resolved simulation despite under-resolution. 

In general, we show that by keeping the kinematic (shear) viscosity coefficient constant the presented method is extremely stable and produces accurate results in presence of under-resolution. These findings and the parameter-free and explicit nature of KBC as well as the lack of explicit turbulence models renders the scheme a very promising candidate for applications in both research and engineering contexts were high Reynolds numbers and computational cost are of importance.

It has been demonstrated in \cite{KBC} that also for three dimensional flows in presence of complex walls low-order statistics can be captured well using KBC. In a further publication we will address the issue of boundary conditions for KBC models in both two and three dimensions.

\section{Acknowledgements}

This work was supported by the European Research Council (ERC) Advanced Grant No. 291094-ELBM. 
Computational resources at the Swiss National Super Computing Center CSCS were provided under the grant S492.

\bibliography{kbc_3d}

\appendix*

\begin{widetext}

\section{Central Moments}

The mapping between natural and central moments is linear in the non-conserved moments and given by the following relations:

\begin{align}\label{eq:cm1}
\begin{split}
\widetilde{\Pi}_{xy} &= \Pi_{xy} - u_x u_y \\
\end{split}
\end{align}
\begin{align}\label{eq:cm2}
\begin{split}
\widetilde{\Pi}_{xz} &= \Pi_{xz} - u_x u_z \\
\end{split}
\end{align}
\begin{align}\label{eq:cm3}
\begin{split}
\widetilde{\Pi}_{yz} &= \Pi_{yz} - u_y u_z \\
\end{split}
\end{align}
\begin{align}\label{eq:cm4}
\begin{split}
\widetilde{N}_{xz} &= N_{xz}-u_x^2+u_z^2 \\
\end{split}
\end{align}
\begin{align}\label{eq:cm5}
\begin{split}
\widetilde{N}_{yz} &= N_{yz}-u_y^2+u_z^2 \\
\end{split}
\end{align}
\begin{align}\label{eq:cm6}
\begin{split}
\widetilde{T} &= T-(u_x^2+u_y^2+u_z^2) \\
\end{split}
\end{align}
\begin{align}\label{eq:cm7}
\begin{split}
\widetilde{Q}_{xyz} &= Q_{xyz} - u_x \widetilde{\Pi}_{yz}- u_y \widetilde{\Pi}_{xz}- u_z\widetilde{\Pi}_{xy} -u_x u_y u_z \\
\end{split}
\end{align}
\begin{align}\label{eq:cm8}
\begin{split}
\widetilde{Q}_{xyy} &= Q_{xyy} -\tfrac{1}{3} \left(6 u_y \widetilde{\Pi}_{xy} + u_x (3 u_y^2 + 2 \widetilde{N}_{yz} - \widetilde{N}_{xz} + \widetilde{T})\right) \\
\end{split}
\end{align}
\begin{align}\label{eq:cm9}
\begin{split}
\widetilde{Q}_{xzz} &= Q_{xzz} -\tfrac{1}{3} \left(6 u_z \widetilde{\Pi}_{xz} + u_x (3 u_z^2 -   \widetilde{N}_{xz} - \widetilde{N}_{yz} + \widetilde{T})\right) \\
\end{split}
\end{align}
\begin{align}\label{eq:cm10}
\begin{split}
\widetilde{Q}_{xxy} &= Q_{xxy} -\tfrac{1}{3} \left(6 u_x \widetilde{\Pi}_{xy} + u_y (3 u_x^2 + 2 \widetilde{N}_{xz} - \widetilde{N}_{yz} + \widetilde{T})\right) \\
\end{split}
\end{align}
\begin{align}\label{eq:cm11}
\begin{split}
\widetilde{Q}_{yzz} &= Q_{yzz} -\tfrac{1}{3} \left(6 u_z \widetilde{\Pi}_{yz} + u_y (3 u_z^2 -   \widetilde{N}_{xz} - \widetilde{N}_{yz} + \widetilde{T})\right) \\
\end{split}
\end{align}
\begin{align}\label{eq:cm12}
\begin{split}
\widetilde{Q}_{xxz} &= Q_{xxz} -\tfrac{1}{3} \left(6 u_x \widetilde{\Pi}_{xz} + u_z (3 u_x^2 + 2 \widetilde{N}_{xz} - \widetilde{N}_{yz} + \widetilde{T})\right) \\
\end{split}
\end{align}
\begin{align}\label{eq:cm13}
\begin{split}
\widetilde{Q}_{yyz} &= Q_{yyz} -\tfrac{1}{3} \left(6 u_y \widetilde{\Pi}_{yz} + u_z (3 u_y^2 + 2 \widetilde{N}_{yz} - \widetilde{N}_{xz} + \widetilde{T})\right) \\
\end{split}
\end{align}
\begin{align}\label{eq:cm14}
\begin{split}
\widetilde{M}_{022} &= {M}_{022} - \left( u_y^2 u_z^2 +                                                                                       4 u_y u_z \widetilde{\Pi}_{yz}                         + (u_y^2+u_z^2) \widetilde{T}/3                   
                           - (u_y^2+u_z^2) \widetilde{N}_{xz}/3                         + (-u_y^2+2 u_z^2) \widetilde{N}_{yz}/3 \right.\\ & \left.
                           +2 u_y \widetilde{Q}_{yzz} + 2 u_z \widetilde{Q}_{yyz} \right) \\
\end{split}
\end{align}
\begin{align}\label{eq:cm15}
\begin{split}
\widetilde{M}_{202} &= {M}_{202} - \left( u_x^2 u_z^2 +                                                4 u_x u_z \widetilde{\Pi}_{xz}                                                                + (u_x^2+u_z^2) \widetilde{T}/3                   
                        + (-u_x^2+2 u_z^2) \widetilde{N}_{xz}/3                            - (u_x^2+u_z^2) \widetilde{N}_{yz}/3 \right.\\ & \left.
                        + 2 u_x \widetilde{Q}_{xzz} + 2 u_z \widetilde{Q}_{xxz}\right) \\
\end{split}
\end{align}
\begin{align}\label{eq:cm16}
\begin{split}
\widetilde{M}_{220} &= {M}_{220} - \left( u_x^2 u_y^2 +         4 u_x u_y \widetilde{\Pi}_{xy}                                                                                                       + (u_x^2+u_y^2) \widetilde{T}/3                   
                        + (-u_x^2+2 u_y^2) \widetilde{N}_{xz}/3                          + (2 u_x^2-u_y^2) \widetilde{N}_{yz}/3 \right.\\ & \left.
                        + 2 u_x \widetilde{Q}_{xyy} + 2 u_y \widetilde{Q}_{xxy}\right) \\
\end{split}
\end{align}
\begin{align}\label{eq:cm17}
\begin{split}
\widetilde{M}_{211} &= {M}_{211} - \left( u_x^2 u_y u_z +         2 u_x u_z \widetilde{\Pi}_{xy}       + 2 u_x u_y \widetilde{\Pi}_{xz}           + u_x^2 \widetilde{\Pi}_{yz}                         +       u_y u_z \widetilde{T}/3                   
                               + 2 u_y u_z \widetilde{N}_{xz}/3                                  - u_y u_z \widetilde{N}_{yz}/3 \right.\\ & \left.
                               +2 u_x \widetilde{Q}_{xyz} + u_z \widetilde{Q}_{xxy} + u_y \widetilde{Q}_{xxz} \right) \\
\end{split}
\end{align}
\begin{align}\label{eq:cm18}
\begin{split}
\widetilde{M}_{121} &= {M}_{121} - \left( u_x u_y^2 u_z +         2 u_y u_z \widetilde{\Pi}_{xy}           + u_y^2 \widetilde{\Pi}_{xz}       + 2 u_x u_y \widetilde{\Pi}_{yz}                         +       u_x u_z \widetilde{T}/3                   
                                 - u_x u_z \widetilde{N}_{xz}/3                                + 2 u_x u_z \widetilde{N}_{yz}/3 \right.\\ & \left.
                                 + 2 u_y \widetilde{Q}_{xyz} + u_z \widetilde{Q}_{xyy} + u_x \widetilde{Q}_{yyz} \right) \\
\end{split}
\end{align}
\begin{align}\label{eq:cm19}
\begin{split}
\widetilde{M}_{112} &= {M}_{112} - \left( u_x u_y u_z^2 +             u_z^2 \widetilde{\Pi}_{xy}       + 2 u_y u_z \widetilde{\Pi}_{xz}       + 2 u_x u_z \widetilde{\Pi}_{yz}                         +       u_x u_y \widetilde{T}/3                   
                                 - u_x u_y \widetilde{N}_{xz}/3                                  - u_x u_y \widetilde{N}_{yz}/3 \right.\\ & \left.
                                 +2 u_z \widetilde{Q}_{xyz} + u_y \widetilde{Q}_{xzz} + u_x \widetilde{Q}_{yzz} \right) \\
\end{split}
\end{align}
\begin{align}\label{eq:cm20}
\begin{split}
\widetilde{M}_{122} &= {M}_{122} - \left( u_x u_y^2 u_z^2 +       2 u_y u_z^2 \widetilde{\Pi}_{xy}     + 2 u_y^2 u_z \widetilde{\Pi}_{xz}   + 4 u_x u_y u_z \widetilde{\Pi}_{yz}                 + (u_x u_y^2+u_x u_z^2) \widetilde{T}/3 \right.\\ & \left.
                  + (-u_x u_y^2-u_x u_z^2) \widetilde{N}_{xz}/3                 + (-u_x u_y^2+2 u_x u_z^2) \widetilde{N}_{yz}/3 \right.\\ & \left.
                  +4 u_y u_z \widetilde{Q}_{xyz} + u_z^2 \widetilde{Q}_{xyy} + u_y^2 \widetilde{Q}_{xzz} + 2 u_x u_y \widetilde{Q}_{yzz} +2 u_x u_z \widetilde{Q}_{yyz} \right.\\ & \left.
                  +2 u_y \widetilde{M}_{112} +2 u_z \widetilde{M}_{121} + u_x \widetilde{M}_{022} \right) \\
\end{split}
\end{align}
\begin{align}\label{eq:cm21}
\begin{split}
\widetilde{M}_{212} &= {M}_{212} - \left( u_x^2 u_y u_z^2 +       2 u_x u_z^2 \widetilde{\Pi}_{xy}   + 4 u_x u_y u_z \widetilde{\Pi}_{xz}     + 2 u_x^2 u_z \widetilde{\Pi}_{yz}                 + (u_x^2 u_y+u_y u_z^2) \widetilde{T}/3 \right.\\ & \left.
                + (-u_x^2 u_y+2 u_y u_z^2) \widetilde{N}_{xz}/3                   + (-u_x^2 u_y-u_y u_z^2) \widetilde{N}_{yz}/3 \right.\\ & \left.
                +4 u_x u_z \widetilde{Q}_{xyz} +2 u_x u_y \widetilde{Q}_{xzz} + u_z^2 \widetilde{Q}_{xxy} + u_x^2 \widetilde{Q}_{yzz} + 2 u_y u_z \widetilde{Q}_{xxz} \right.\\ & \left.
                +2 u_z \widetilde{M}_{211} + 2 u_x \widetilde{M}_{112} + u_y \widetilde{M}_{202} \right) \\
\end{split}
\end{align}
\begin{align}\label{eq:cm22}
\begin{split}
\widetilde{M}_{221} &= {M}_{221} - \left( u_x^2 u_y^2 u_z +     4 u_x u_y u_z \widetilde{\Pi}_{xy}    + 2  u_x u_y^2 \widetilde{\Pi}_{xz}     + 2 u_x^2 u_y \widetilde{\Pi}_{yz}                 + (u_x^2 u_z+u_y^2 u_z) \widetilde{T}/3 \right.\\ & \left.
                + (-u_x^2 u_z+2 u_y^2 u_z) \widetilde{N}_{xz}/3                  + (2 u_x^2 u_z-u_y^2 u_z) \widetilde{N}_{yz}/3 \right.\\ & \left.
                +4 u_x u_y \widetilde{Q}_{xyz} +2 u_x u_z \widetilde{Q}_{xyy} +2 u_y u_z \widetilde{Q}_{xxy} + u_y^2 \widetilde{Q}_{xxz} + u_x^2 \widetilde{Q}_{yyz} \right.\\ & \left.
                +2 u_y \widetilde{M}_{211} + 2 u_x \widetilde{M}_{121} + u_z \widetilde{M}_{220} \right) \\
\end{split}
\end{align}
\begin{align}\label{eq:cm23}
\begin{split}
\widetilde{M}_{222} &= {M}_{222} - \left( u_x^2 u_y^2 u_z^2 +   4 u_x u_y u_z^2 \widetilde{\Pi}_{xy} + 4 u_x u_y^2 u_z \widetilde{\Pi}_{xz} + 4 u_x^2 u_y u_z \widetilde{\Pi}_{yz} + (u_x^2 u_y^2+u_x^2 u_z^2+u_y^2 u_z^2) \widetilde{T}/3 \right.\\ & \left.
+ (-u_x^2 u_y^2-u_x^2 u_z^2+2 u_y^2 u_z^2) \widetilde{N}_{xz}/3 + (-u_x^2 u_y^2+2 u_x^2 u_z^2-u_y^2 u_z^2) \widetilde{N}_{yz}/3 \right.\\ & \left.
+8 u_x u_y u_z \widetilde{Q}_{xyz} + 2 u_x u_z^2 \widetilde{Q}_{xyy} + 2 u_x u_y^2 \widetilde{Q}_{xzz} + 2 u_y u_z^2 \widetilde{Q}_{xxy} + 2 u_x^2 u_y \widetilde{Q}_{yzz} + 2 u_y^2 u_z \widetilde{Q}_{xxz} + 2 u_x^2 u_z \widetilde{Q}_{yyz}\right.\\ & \left.
+u_x^2 \widetilde{M}_{022} + u_y^2 \widetilde{M}_{202} + u_z^2 \widetilde{M}_{220} + 4 u_y u_z \widetilde{M}_{211} + 4 u_x u_z \widetilde{M}_{121} + 4 u_x u_y \widetilde{M}_{112} + 2 u_x \widetilde{M}_{122} + 2 u_y \widetilde{M}_{212} + 2 u_z \widetilde{M}_{221} \right).
\end{split}
\end{align}

Substituting Eqs.~\eqref{eq:cm1}-\eqref{eq:cm23} in Eq.~\eqref{eq:moment_rep} one arrives at the moment representation in the central basis:

\begin{align}\label{eq:cmf000}
\begin{split}
f_{(0,0,0)} &= \rho \left( 
 - \left(u_x^2-1\right) \left(u_y^2-1\right) \left(u_z^2-1\right) \right.\\ & \left. 
 - 4 u_x u_y \left(u_z^2-1\right) \widetilde{\Pi}_{xy}
 - 4 u_x u_z \left(u_y^2-1\right) \widetilde{\Pi}_{xz}
 - 4 u_y u_z \left(u_x^2-1\right) \widetilde{\Pi}_{yz} \right.\\ & \left. 
 + \left( u_x^2\left(u_y^2+u_z^2-2\right) + u_z^2\left(1-2u_y^2\right) + u_y^2 \right) \widetilde{N}_{xz}/3
 + \left( u_y^2\left(u_x^2+u_z^2-2\right) + u_z^2\left(1-2u_x^2\right) + u_x^2 \right) \widetilde{N}_{yz}/3 \right.\\ & \left. 
 - \left(u_x^2 u_y^2 + u_x^2 u_z^2 + u_y^2 u_z^2 - 2 \left( u_x^2 + u_y^2 + u_z^2 \right) + 3\right) \widetilde{T}/3 \right.\\ & \left. 
 - 8 u_x u_y u_z \widetilde{Q}_{xyz} \right.\\ & \left. 
 - 2 u_x \left(u_z^2-1\right) \widetilde{Q}_{xyy}
 - 2 u_x \left(u_y^2-1\right) \widetilde{Q}_{xzz} \right.\\ & \left. 
 - 2 u_y \left(u_z^2-1\right) \widetilde{Q}_{xxy}
 - 2 u_y \left(u_x^2-1\right) \widetilde{Q}_{yzz} \right.\\ & \left. 
 - 2 u_z \left(u_y^2-1\right) \widetilde{Q}_{xxz}
 - 2 u_z \left(u_x^2-1\right) \widetilde{Q}_{yyz} \right.\\ & \left. 
 + \left(1-u_x^2\right) \widetilde{M}_{022}
 + \left(1-u_y^2\right) \widetilde{M}_{202}
 + \left(1-u_z^2\right) \widetilde{M}_{220} \right.\\ & \left. 
 - 4 u_y u_z \widetilde{M}_{211}
 - 4 u_x u_z \widetilde{M}_{121}
 - 4 u_x u_y \widetilde{M}_{112} \right.\\ & \left. 
 - 2 u_x \widetilde{M}_{122}
 - 2 u_y \widetilde{M}_{212}
 - 2 u_z \widetilde{M}_{221} 
 - \widetilde{M}_{222}
\right ) \\
\end{split}
\end{align}
\begin{align}\label{eq:cmfx00}
\begin{split}
f_{(\sigma,0,0)} &= \tfrac{1}{6} \rho \left( 
3 u_x \left(u_y^2-1\right) \left(u_z^2-1\right) (\sigma +u_x) \right.\\ & \left. 
 + 6 u_y \left(u_z^2-1\right) (\sigma +2 u_x) \widetilde{\Pi}_{xy}
 + 6 u_z \left(u_y^2-1\right) (\sigma +2 u_x) \widetilde{\Pi}_{xz}
 + 12 u_x u_y u_z (\sigma +u_x) \widetilde{\Pi}_{yz} \right.\\ & \left. 
 - \left(\sigma  u_x \left(u_y^2+u_z^2-2\right)+u_x^2 \left(u_y^2+u_z^2-2\right)-2 \left(u_y^2-1\right) \left(u_z^2-1\right)\right) \widetilde{N}_{xz} \right.\\ & \left. 
  +\left(u_z^2 \left(2 u_x (\sigma +u_x)-u_y^2\right)-u_x \left(u_y^2+1\right) (\sigma +u_x)+u_y^2+u_z^2-1\right) \widetilde{N}_{yz} \right.\\ & \left. 
 + \left(u_z^2 \left(u_x (\sigma +u_x)+u_y^2\right)+u_x \left(u_y^2-2\right) (\sigma +u_x)-u_y^2-u_z^2+1\right) \widetilde{T} \right.\\ & \left. 
 + 12 u_y u_z (\sigma +2 u_x) \widetilde{Q}_{xyz} \right.\\ & \left. 
 + 3\left(u_z^2-1\right) (\sigma +2 u_x) \widetilde{Q}_{xyy}
 + 3\left(u_y^2-1\right) (\sigma +2 u_x) \widetilde{Q}_{xzz} \right.\\ & \left. 
 + 6 u_y \left(u_z^2-1\right) \widetilde{Q}_{xxy}
 + 6 u_x u_y (\sigma +u_x) \widetilde{Q}_{yzz} \right.\\ & \left. 
 + 6 u_z \left(u_y^2-1\right) \widetilde{Q}_{xxz}
 + 6 u_x u_z (\sigma +u_x) \widetilde{Q}_{yyz} \right.\\ & \left. 
 + 3 u_x (\sigma +u_x) \widetilde{M}_{022}
 + 3 \left(u_y^2-1\right) \widetilde{M}_{202}
 + 3 \left(u_z^2-1\right) \widetilde{M}_{220} \right.\\ & \left. 
 + 12 u_y u_z \widetilde{M}_{211}
 + 6 u_z (\sigma +2 u_x) \widetilde{M}_{121}
 + 6 u_y (\sigma +2 u_x) \widetilde{M}_{112} \right.\\ & \left. 
 + 3 \left(\sigma +2 u_x\right) \widetilde{M}_{122}
 + 6 u_y \widetilde{M}_{212}
 + 6 u_z \widetilde{M}_{221} 
 + 3 \widetilde{M}_{222}
\right ) \\
\end{split}
\end{align}
\begin{align}\label{eq:cmf0y0}
\begin{split}
f_{(0,\lambda,0)} &= \tfrac{1}{6} \rho \left( 
 3 u_y \left(u_x^2-1\right) \left(u_z^2-1\right) (\lambda +u_y) \right.\\ & \left. 
 + 6 u_x \left(u_z^2-1\right) (\lambda +2 u_y) \widetilde{\Pi}_{xy}
 + 12  u_x u_y u_z (\lambda +u_y) \widetilde{\Pi}_{xz}
 + 6 u_z \left(u_x^2-1\right) (\lambda +2 u_y) \widetilde{\Pi}_{yz} \right.\\ & \left. 
 + \left(-u_x^2 \left(u_y (\lambda +u_y)+u_z^2-1\right)+u_y \left(2 u_z^2-1\right) (\lambda +u_y)+u_z^2-1\right) \widetilde{N}_{xz} \right.\\ & \left. 
 + \left(-u_x^2 \left(u_y (\lambda +u_y)-2 u_z^2+2\right)-u_y \left(u_z^2-2\right) (\lambda +u_y)-2 u_z^2+2\right) \widetilde{N}_{yz} \right.\\ & \left. 
 + \left(u_x^2 \left(u_y (\lambda +u_y)+u_z^2-1\right)+u_y \left(u_z^2-2\right) (\lambda +u_y)-u_z^2+1\right) \widetilde{T} \right.\\ & \left. 
 + 12 u_x u_z (\lambda +2 u_y) \widetilde{Q}_{xyz} \right.\\ & \left. 
 + 6 u_x \left(u_z^2-1\right) \widetilde{Q}_{xyy}
 + 6 u_x u_y (\lambda +u_y) \widetilde{Q}_{xzz} \right.\\ & \left. 
 + 3 \left(u_z^2-1\right) (\lambda +2 u_y) \widetilde{Q}_{xxy}
 + 3 \left(u_x^2-1\right) (\lambda +2 u_y) \widetilde{Q}_{yzz} \right.\\ & \left. 
 + 6 u_y u_z (\lambda +u_y) \widetilde{Q}_{xxz}
 + 6 u_z \left(u_x^2-1\right) \widetilde{Q}_{yyz} \right.\\ & \left. 
 + 3 \left(u_x^2-1\right) \widetilde{M}_{022}
 + 3 u_y (\lambda +u_y) \widetilde{M}_{202}
 + 3 \left(u_z^2-1\right) \widetilde{M}_{220} \right.\\ & \left. 
 + 6 u_z (\lambda +2 u_y) \widetilde{M}_{211}
 + 12 u_x u_z \widetilde{M}_{121}
 + 6 u_x (\lambda +2 u_y) \widetilde{M}_{112} \right.\\ & \left. 
 + 6 u_x \widetilde{M}_{122}
 + 3 \left(\lambda +2 u_y\right) \widetilde{M}_{212}
 + 6 u_z \widetilde{M}_{221} 
 + 3 \widetilde{M}_{222}
\right ) \\
\end{split}
\end{align}
\begin{align}\label{eq:cmf00z}
\begin{split}
f_{(0,0,\delta)} &= \tfrac{1}{6} \rho \left( 
  3 u_z \left(u_x^2-1\right) \left(u_y^2-1\right) (\delta +u_z) \right.\\ & \left. 
 + 12 u_x u_y u_z (\delta +u_z) \widetilde{\Pi}_{xy}
 + 6 u_x \left(u_y^2-1\right) (\delta +2 u_z) \widetilde{\Pi}_{xz}
 + 6 u_y \left(u_x^2-1\right) (\delta +2 u_z) \widetilde{\Pi}_{yz} \right.\\ & \left. 
 + \left(-u_x^2 \left(u_z (\delta +u_z)+u_y^2-1\right)+u_y^2 (2 u_z (\delta +u_z)+1)-u_z (\delta +u_z)-1\right) \widetilde{N}_{xz} \right.\\ & \left. 
 + \left(u_x^2 \left(2 u_z (\delta +u_z)-u_y^2+1\right)-u_y^2 (u_z (\delta +u_z)-1)-u_z (\delta +u_z)-1\right) \widetilde{N}_{yz} \right.\\ & \left. 
 + \left(u_x^2 \left(u_z (\delta +u_z)+u_y^2-1\right)+u_y^2 (u_z (\delta +u_z)-1)-2 u_z (\delta +u_z)+1\right) \widetilde{T} \right.\\ & \left. 
 + 12 u_x u_y (\delta +2 u_z) \widetilde{Q}_{xyz} \right.\\ & \left. 
 + 6 u_x u_z (\delta +u_z) \widetilde{Q}_{xyy}
 + 6 u_x \left(u_y^2-1\right) \widetilde{Q}_{xzz} \right.\\ & \left. 
 + 6 u_y u_z (\delta +u_z) \widetilde{Q}_{xxy}
 + 6 u_y \left(u_x^2-1\right) \widetilde{Q}_{yzz} \right.\\ & \left. 
 + 3 \left(u_y^2-1\right) (\delta +2 u_z) \widetilde{Q}_{xxz}
 + 3 \left(u_x^2-1\right) (\delta +2 u_z) \widetilde{Q}_{yyz} \right.\\ & \left. 
 + 3 \left(u_x^2-1\right) \widetilde{M}_{022}
 + 3 \left(u_y^2-1\right) \widetilde{M}_{202}
 + 3 u_z (\delta +u_z) \widetilde{M}_{220} \right.\\ & \left. 
 + 6 u_y (\delta +2 u_z) \widetilde{M}_{211}
 + 6 u_x (\delta +2 u_z) \widetilde{M}_{121}
 + 12 u_x u_y \widetilde{M}_{112} \right.\\ & \left. 
 + 6 u_x \widetilde{M}_{122}
 + 6 u_y \widetilde{M}_{212}
 + 3 \left(\delta +2 u_z\right) \widetilde{M}_{221} 
 + 3 \widetilde{M}_{222}
\right ) \\
\end{split}
\end{align}
\begin{align}\label{eq:cmfxy0}
\begin{split}
f_{(\sigma,\lambda,0)} &= \tfrac{1}{4} \rho \left( 
 - u_x u_y \left(u_z^2-1\right) (\lambda +u_y) (\sigma +u_x) \right.\\ & \left. 
 - \left(u_z^2-1\right) (\lambda +2 u_y) (\sigma +2 u_x) \widetilde{\Pi}_{xy}
 -2 u_y u_z (\lambda +u_y) (\sigma +2 u_x) \widetilde{\Pi}_{xz}
 - u_x u_z (\lambda +2 u_y) (\sigma +u_x) \widetilde{\Pi}_{yz} \right.\\ & \left. 
 + \tfrac{1}{3} \left(\sigma  u_x \left(u_y (\lambda +u_y)+u_z^2-1\right)+u_x^2 \left(u_y (\lambda +u_y)+u_z^2-1\right)-2 u_y \left(u_z^2-1\right) (\lambda +u_y)\right) \widetilde{N}_{xz} \right.\\ & \left. 
 + \tfrac{1}{3} \left(\sigma  u_x \left(u_y (\lambda +u_y)-2 u_z^2+2\right)+u_x^2 \left(u_y (\lambda +u_y)-2 u_z^2+2\right)+u_y \left(u_z^2-1\right) (\lambda +u_y)\right) \widetilde{N}_{yz} \right.\\ & \left. 
 + \tfrac{1}{3} \left(-\sigma  u_x \left(u_y (\lambda +u_y)+u_z^2-1\right)-u_x^2 \left(u_y (\lambda +u_y)+u_z^2-1\right)-u_y \left(u_z^2-1\right) (\lambda +u_y)\right) \widetilde{T} \right.\\ & \left. 
 - 2 u_z (\lambda +2 u_y) (\sigma +2 u_x) \widetilde{Q}_{xyz} \right.\\ & \left. 
 - \left(u_z^2-1\right) (\sigma +2 u_x) \widetilde{Q}_{xyy}
 - u_y (\lambda +u_y) (\sigma +2 u_x) \widetilde{Q}_{xzz} \right.\\ & \left. 
 - \left(u_z^2-1\right) (\lambda +2 u_y) \widetilde{Q}_{xxy}
 - u_x (\lambda +2 u_y) (\sigma +u_x) \widetilde{Q}_{yzz} \right.\\ & \left. 
 - 2 u_y u_z (\lambda +u_y) \widetilde{Q}_{xxz}
 - 2 u_x u_z (\sigma +u_x) \widetilde{Q}_{yyz} \right.\\ & \left. 
 - u_x (\sigma +u_x) \widetilde{M}_{022}
 - u_y (\lambda +u_y) \widetilde{M}_{202}
 + \left(1-u_z^2\right) \widetilde{M}_{220} \right.\\ & \left. 
 - 2 u_z (\lambda +2 u_y) \widetilde{M}_{211}
 - 2 u_z (\sigma +2 u_x) \widetilde{M}_{121}
 - (\lambda +2 u_y) (\sigma +2 u_x) \widetilde{M}_{112} \right.\\ & \left. 
 - (\sigma +2 u_x) \widetilde{M}_{122}
 - (\lambda +2 u_y) \widetilde{M}_{212}
 - 2 u_z \widetilde{M}_{221} 
 - \widetilde{M}_{222}
\right ) \\
\end{split}
\end{align}
\begin{align}\label{eq:cmfx0z}
\begin{split}
f_{(\sigma,0,\delta)} &= \tfrac{1}{4} \rho \left( 
 - u_x u_z \left(u_y^2-1\right) (\delta +u_z) (\sigma +u_x) \right.\\ & \left. 
 - 2 u_y u_z (\delta +u_z) (\sigma +2 u_x) \widetilde{\Pi}_{xy}
 - \left(u_y^2-1\right) (\delta +2 u_z) (\sigma +2 u_x) \widetilde{\Pi}_{xz}
 - 2 u_x u_y (\delta +2 u_z) (\sigma +u_x) \widetilde{\Pi}_{yz} \right.\\ & \left. 
 + \tfrac{1}{3} \left(\sigma  u_x \left(u_z (\delta +u_z)+u_y^2-1\right)+u_x^2 \left(u_z (\delta +u_z)+u_y^2-1\right)-2 \left(u_y^2-1\right) u_z (\delta +u_z)\right) \widetilde{N}_{xz} \right.\\ & \left. 
 + \tfrac{1}{3} \left(\sigma  u_x \left(-2 u_z (\delta +u_z)+u_y^2-1\right)+u_x^2 \left(-2 u_z (\delta +u_z)+u_y^2-1\right)+\left(u_y^2-1\right) u_z (\delta +u_z)\right) \widetilde{N}_{yz} \right.\\ & \left. 
 + \tfrac{1}{3} \left(-\sigma  u_x \left(u_z (\delta +u_z)+u_y^2-1\right)-u_x^2 \left(u_z (\delta +u_z)+u_y^2-1\right)-\left(u_y^2-1\right) u_z (\delta +u_z)\right) \widetilde{T} \right.\\ & \left. 
 - 2 u_y (\delta +2 u_z) (\sigma +2 u_x) \widetilde{Q}_{xyz} \right.\\ & \left. 
 - u_z (\delta +u_z) (\sigma +2 u_x) \widetilde{Q}_{xyy}
 - \left(u_y^2-1\right) (\sigma +2 u_x) \widetilde{Q}_{xzz} \right.\\ & \left. 
 - 2 u_y u_z (\delta +u_z) \widetilde{Q}_{xxy}
 - 2 u_x u_y (\sigma +u_x) \widetilde{Q}_{yzz} \right.\\ & \left. 
 - \left(u_y^2-1\right) (\delta +2 u_z) \widetilde{Q}_{xxz}
 - u_x (\delta +2 u_z) (\sigma +u_x) \widetilde{Q}_{yyz} \right.\\ & \left. 
 - u_x (\sigma +u_x) \widetilde{M}_{022}
 + \left(1-u_y^2\right) \widetilde{M}_{202}
 - u_z (\delta +u_z) \widetilde{M}_{220} \right.\\ & \left. 
 - 2 u_y (\delta +2 u_z) \widetilde{M}_{211}
 - (\delta +2 u_z) (\sigma +2 u_x) \widetilde{M}_{121}
 - 2 u_y (\sigma +2 u_x) \widetilde{M}_{112} \right.\\ & \left. 
 - (\sigma +2 u_x) \widetilde{M}_{122}
 - 2 u_y \widetilde{M}_{212}
 - (\delta +2 u_z) \widetilde{M}_{221} 
 - \widetilde{M}_{222}
\right ) \\
\end{split}
\end{align}
\begin{align}\label{eq:cmf0yz}
\begin{split}
f_{(0,\lambda,\delta)} &= \tfrac{1}{4} \rho \left( 
 - u_y u_z \left(u_x^2-1\right) (\delta +u_z) (\lambda +u_y) \right.\\ & \left. 
 - 2 u_x u_z (\delta +u_z) (\lambda +2 u_y) \widetilde{\Pi}_{xy}
 - 2 u_x u_y (\delta +2 u_z) (\lambda +u_y) \widetilde{\Pi}_{xz}
 - \left(u_x^2-1\right) (\delta +2 u_z) (\lambda +2 u_y) \widetilde{\Pi}_{yz} \right.\\ & \left. 
 + \tfrac{1}{3} \left(\lambda  u_y \left(-2 u_z (\delta +u_z)+u_x^2-1\right)+u_y^2 \left(-2 u_z (\delta +u_z)+u_x^2-1\right)+\left(u_x^2-1\right) u_z (\delta +u_z)\right) \widetilde{N}_{xz} \right.\\ & \left. 
 + \tfrac{1}{3} \left(\lambda  u_y \left(u_z (\delta +u_z)+u_x^2-1\right)+u_y^2 \left(u_z (\delta +u_z)+u_x^2-1\right)-2 \left(u_x^2-1\right) u_z (\delta +u_z)\right) \widetilde{N}_{yz} \right.\\ & \left. 
 + \tfrac{1}{3} \left(-\lambda  u_y \left(u_z (\delta +u_z)+u_x^2-1\right)-u_y^2 \left(u_z (\delta +u_z)+u_x^2-1\right)-\left(u_x^2-1\right) u_z (\delta +u_z)\right) \widetilde{T} \right.\\ & \left. 
 - 2 u_x (\delta +2 u_z) (\lambda +2 u_y) \widetilde{Q}_{xyz} \right.\\ & \left. 
 - 2 u_x u_z (\delta +u_z) \widetilde{Q}_{xyy}
 - 2 u_x u_y (\lambda +u_y) \widetilde{Q}_{xzz} \right.\\ & \left. 
 - u_z (\delta +u_z) (\lambda +2 u_y) \widetilde{Q}_{xxy}
 - \left(u_x^2-1\right) (\lambda +2 u_y) \widetilde{Q}_{yzz} \right.\\ & \left. 
 - u_y (\delta +2 u_z) (\lambda +u_y) \widetilde{Q}_{xxz}
 - \left(u_x^2-1\right) (\delta +2 u_z) \widetilde{Q}_{yyz} \right.\\ & \left. 
 + \left(1-u_x^2\right) \widetilde{M}_{022}
 - u_y (\lambda +u_y) \widetilde{M}_{202}
 - u_z (\delta +u_z)\widetilde{M}_{220} \right.\\ & \left. 
 - (\delta +2 u_z) (\lambda +2 u_y) \widetilde{M}_{211}
 - 2 u_x (\delta +2 u_z) \widetilde{M}_{121}
 - 2 u_x (\lambda +2 u_y) \widetilde{M}_{112} \right.\\ & \left. 
 - 2 u_x \widetilde{M}_{122}
 - (\lambda +2 u_y) \widetilde{M}_{212}
 - (\delta +2 u_z) \widetilde{M}_{221} 
 - \widetilde{M}_{222}
\right ) \\
\end{split}
\end{align}
\begin{align}\label{eq:cmfxyz}
\begin{split}
f_{(\sigma,\lambda,\delta)} &= \tfrac{1}{8} \rho \left( 
  u_x u_y u_z (\delta +u_z) (\lambda +u_y) (\sigma +u_x) \right.\\ & \left. 
 + u_z (\delta +u_z) (\lambda +2 u_y) (\sigma +2 u_x) \widetilde{\Pi}_{xy}
 + u_y (\delta +2 u_z) (\lambda +u_y) (\sigma +2 u_x) \widetilde{\Pi}_{xz}
 + u_x (\delta +2 u_z) (\lambda +2 u_y) (\sigma +u_x) \widetilde{\Pi}_{yz} \right.\\ & \left. 
 + \tfrac{1}{3} \left(-\delta  u_z (u_x (\sigma +u_x)-2 u_y (\lambda +u_y))-u_z^2 (u_x (\sigma +u_x)-2 u_y (\lambda +u_y))-u_x u_y (\lambda +u_y) (\sigma +u_x)\right) \widetilde{N}_{xz} \right.\\ & \left. 
 + \tfrac{1}{3} \left(\delta  u_z (2 u_x (\sigma +u_x)-u_y (\lambda +u_y))+u_z^2 (2 u_x (\sigma +u_x)-u_y (\lambda +u_y))-u_x u_y (\lambda +u_y) (\sigma +u_x)\right) \widetilde{N}_{yz} \right.\\ & \left. 
 + \tfrac{1}{3} \left(\delta  u_z (u_y (\lambda +u_y)+u_x (\sigma +u_x))+u_z^2 (u_y (\lambda +u_y)+u_x (\sigma +u_x))+u_x u_y (\lambda +u_y) (\sigma +u_x)\right) \widetilde{T} \right.\\ & \left. 
 + (\delta +2 u_z) (\lambda +2 u_y) (\sigma +2 u_x) \widetilde{Q}_{xyz} \right.\\ & \left. 
 + u_z (\delta +u_z) (\sigma +2 u_x) \widetilde{Q}_{xyy}
 + u_y (\lambda +u_y) (\sigma +2 u_x) \widetilde{Q}_{xzz} \right.\\ & \left. 
 + u_z (\delta +u_z) (\lambda +2 u_y) \widetilde{Q}_{xxy}
 + u_x (\lambda +2 u_y) (\sigma +u_x) \widetilde{Q}_{yzz} \right.\\ & \left. 
 + u_y (\delta +2 u_z) (\lambda +u_y) \widetilde{Q}_{xxz}
 + u_x (\delta +2 u_z) (\sigma +u_x) \widetilde{Q}_{yyz} \right.\\ & \left. 
 + u_x (\sigma +u_x) \widetilde{M}_{022}
 + u_y (\lambda +u_y) \widetilde{M}_{202}
 + u_z (\delta +u_z) \widetilde{M}_{220} \right.\\ & \left. 
 + (\delta +2 u_z) (\lambda +2 u_y) \widetilde{M}_{211}
 + (\delta +2 u_z) (\sigma +2 u_x) \widetilde{M}_{121}
 + (\lambda +2 u_y) (\sigma +2 u_x) \widetilde{M}_{112} \right.\\ & \left. 
 + (\sigma +2 u_x) \widetilde{M}_{122}
 + (\lambda +2 u_y) \widetilde{M}_{212}
 + (\delta +2 u_z) \widetilde{M}_{221} 
 + \widetilde{M}_{222}
\right ) \\
\end{split}
\end{align}
\end{widetext}

\end{document}